\documentclass[journal]{IEEEtran}
\usepackage[T1]{fontenc}
\usepackage{amsmath,amsfonts,amssymb}
\usepackage[thmmarks,amsmath]{ntheorem}
\usepackage{cases}
\usepackage{graphicx}
\usepackage{fixfoot}
\usepackage{stfloats,balance}
\usepackage{epsfig}
\usepackage[noend]{algpseudocode}
\usepackage{algorithmicx,algorithm}
\usepackage{cite}
\usepackage{xcolor}
\usepackage{subeqnarray}
\usepackage{textcomp}
\usepackage{bm}
\theoremseparator{:}
\theoremheaderfont{\itshape}

\newtheorem{theorem}{Theorem}

\newtheorem{remark}{Remark}
\newtheorem{property}{Property}
\allowdisplaybreaks[3]
\def\BibTeX{{\rm B\kern-.05em{\sc i\kern-.025em b}\kern-.08em
    T\kern-.1667em\lower.7ex\hbox{E}\kern-.125emX}}
    
\allowdisplaybreaks[4] 
\begin{document}
\title{ Cell-Free Satellite-UAV Networks for 6G Wide-Area Internet of Things }
 \author{Chengxiao~Liu,
         Wei~Feng,~\IEEEmembership{Senior Member,~IEEE,}
        Yunfei~Chen,~\IEEEmembership{Senior~Member,~IEEE,}
       Cheng-Xiang~Wang,~\IEEEmembership{Fellow,~IEEE,}
       and Ning Ge,~\IEEEmembership{Member,~IEEE}
\thanks{
Manuscript received February 01, 2020; revised June 10, 2020; accepted July 17, 2020. This work was supported in part by the National Key R\&D Program of China under Grant 2018YFA0701601, the National Natural Science Foundation of China (Grant No. 61922049, 61771286, 61941104, 61960206006, 61701457, 91638205), the Frontiers Science Center for Mobile Information Communication and Security, the High Level Innovation and Entrepreneurial Research Team Program in Jiangsu, the High Level Innovation and Entrepreneurial Talent Introduction Program in Jiangsu, the Research Fund of National Mobile Communications Research Laboratory, Southeast University, under Grant 2020B01, the Fundamental Research Funds for the Central Universities under Grant 2242020R30001, the Huawei Cooperation Project, the EU H2020 RISE TESTBED2 project under Grant 872172, the Nantong Technology Program under Grant JC2019115, and the Beijing Innovation Center for Future Chip. This work was presented in part at the IEEE WOCC'2020 \cite{Liu2020WOCC}.}
\thanks{C. Liu, W. Feng (corresponding author), and N. Ge are with the Beijing
National Research Center for Information Science and Technology, Department of Electronic Engineering, Tsinghua University, Beijing 100084, China, W. Feng is also with the Peng Cheng Laboratory, Shenzhen 518055, China (email: lcx17@mails.tsinghua.edu.cn, fengwei@tsinghua.edu.cn, gening@tsinghua.edu.cn).}
\thanks{Y. Chen is with the School of Engineering, University of Warwick, Coventry
CV4 7AL, U.K. (e-mail: Yunfei.Chen@warwick.ac.uk).}
\thanks{C.-X. Wang is with the National Mobile Communications Research Laboratory,
School of Information Science and Engineering, Southeast University,
Nanjing 210096, China, and also with Purple Mountain Laboratories, Nanjing
211111, China (e-mail: chxwang@seu.edu.cn).}
}
\maketitle
\begin{abstract}
In fifth generation (5G) and beyond Internet of Things (IoT), it becomes increasingly important to serve a massive number of IoT devices outside the coverage of terrestrial cellular networks. Due to their own limitations, unmanned aerial vehicles
(UAVs) and satellites need to coordinate with each other in the coverage holes of 5G, leading to a cognitive satellite-UAV network (CSUN). In this paper, we investigate multi-domain resource allocation for CSUNs consisting of a satellite and a swarm of UAVs, so as to improve the efficiency of massive access in wide areas. Particularly, the cell-free on-demand coverage is established to overcome the cost-ineffectiveness of conventional cellular architecture. Opportunistic spectrum sharing is also implemented to cope with the spectrum scarcity problem. To this end, a process-oriented optimization framework is proposed for jointly allocating subchannels, transmit power and hovering times, which considers the whole flight process of UAVs and uses only the slowly-varying large-scale channel state information (CSI). Under the on-board energy constraints of UAVs and interference temperature constraints from UAV swarm to satellite users, we present iterative multi-domain resource allocation algorithms to improve network efficiency with guaranteed user fairness. Simulation results demonstrate the superiority of the proposed algorithms. Moreover, the adaptive cell-free coverage pattern is observed, which implies a promising way to efficiently serve wide-area IoT devices in the upcoming sixth generation (6G) era. 
\end{abstract}
\section{Introduction}
\par{
In fifth generation (5G) and upcoming sixth generation (6G) networks, the demand for wide-area Internet of Things (IoT) with a massive number of devices keeps increasing \cite{Jia2020, Qi2020, Ikpehai2019}. Thus, it is critical to support massive access in emerging terrestrial and satellite networks \cite{Zhen2019}. However, limited by geographical environments, most IoT devices, e.g., buoys on the ocean and sensors in the remote area, are outside the coverage of terrestrial cellular networks \cite{Ikpehai2019}. Consequently, it is hard for conventional IoT technologies, such as Narrow Band IoT (NB-IoT) and Long Range Radio (LoRa), to be used for wide-area IoT directly. Besides, it is also challenging for current satellite networks to serve these devices, due to their limited communication rate and inherent large latency \cite{Zhen2019}.
}
\par{
To overcome these challenges, it is widely regarded as an effective way to integrate unmanned aerial vehicles (UAVs) with satellite networks. Nevertheless, new difficulties arise in building a hybrid satellite-UAV network to efficiently support massive access for wide-area IoT. For example, IoT devices are always sparsely and unevenly distributed in wide areas \cite{Yang2018,Baek2019}, so that it is cost-ineffective to cover them using conventional cellular architecture \cite{Lyu2018}. Furthermore, the spectrum scarcity problem becomes serious, because local spectrum reuse as cellular architecture is no longer applicable \cite{Pandian2015}, due to the mobility of UAVs and ubiquitous coverage of satellites. Thus, opportunistic spectrum sharing for satellite-UAV networks requires global optimization to tackle the wide-area coupled interference. To solve these problems, we investigate the wide-area IoT-oriented cell-free cognitive satellite-UAV network (CSUN), which remains open to our knowledge. }
\subsection{Related Works}
NB-IoT is a widely-used IoT technology for massive connectivity~\cite{Raza2017,Popli2019}, which has been shown effective in urban areas \cite{Raza2017}. However, NB-IoT was designed based on conventional cellular architectures. As shown in \cite{Onireti2016}, the cellular architecture is expensive for bringing services to rural areas. When the IoT devices are sparsely deployed within a vast area, the efficiency of NB-IoT will degrade. Moreover, it is difficult to establish an NB-IoT network on the ocean or in a mountainous area, where the deployment of communication infrastructures is quite limited due to geographical conditions. Likewise, LoRa, as another promising IoT technology~\cite{Centenaro2016,Raza2017} which can serve IoT devices up to tens of kilometers away from the gateway \cite{Ikpehai2019}, also faces challenges in harsh deploying environments. 

To serve a massive number of IoT devices outside the coverage of terrestrial networks,
satellite is widely regarded as a promising enabler \cite{Sanctis2016, Wang2018}. The authors of \cite{Sanctis2016} discussed the group-based massive connectivity for satellite-enabled IoT networks, where spectrum efficiency is shown to be a huge bottleneck. In \cite{Wang2018}, a non-orthogonal slotted Aloha based multiple access framework was proposed for satellites, which can support massive access with narrow bandwidth at the expense of large latency. These works have shown that limited spectrum, lack of efficiency and large latency are main challenges for satellite-enabled IoT networks. To handle the spectrum scarcity problem, cognitive spectrum sharing techniques can be used \cite{Maleki2015}, for which interference mitigation techniques are crucial. In \cite{Vazquez2018}, a hybrid analog-digital transmit beamforming scheme was proposed to mitigate the satellite-terrestrial interference. The authors of \cite{Khan2012} proposed a semi-adaptive beamforming scheme for hybrid satellite-terrestrial networks. In \cite{Liu2019Optimal}, an optimal beamforming method was designed considering nonlinear power amplifiers and imperfect channel knowledge. However, these techniques mainly focused on the spectrum sharing between satellites and fixed terrestrial networks. Due to the mobility of UAVs, cognitive spectrum sharing should be redesigned for CSUNs, where the spatial distribution of interference is much more dynamic. 

\par{
Indeed, UAV is another choice to support massive access for IoT devices \cite{Feng2018UAV, Zeng2019, Salam2019,Hattab2020,Bushnaq2019}. In \cite{Zeng2019}, a whole and worthwhile picture of UAV-enabled 5G and beyond networks was comprehensively investigated. The authors of \cite{Salam2019} discussed the energy efficiency of data aggregation in UAV-enabled IoT networks. In \cite{Hattab2020}, UAVs and cellular networks shared spectrum to improve the performance of data aggregation, where the energy efficiency of IoT devices was also maximized. The authors of \cite{Bushnaq2019} optimized the total flight time of UAVs to save energy through path planning with guaranteed data aggregation efficiency. However, limited on-board energy and limited coverage of a single UAV are still challenging obstacles, which motivate the utilization of UAV swarm and the integration of UAVs with satellites \cite{Wu2018, Khuwaja2019, Hua2019, Wang2019}. In \cite{Wu2018}, the user scheduling and association, transmit power and trajectory of UAV swarm were elaborately optimized in a joint way to improve the worst-case performance of UAVs. The authors of \cite{Khuwaja2019} studied the placement of a swarm of UAVs to optimize the coverage area with co-channel interference. In \cite{Hua2019}, a coordinated multi-point transmission scheme was proposed for a UAV-aided cognitive satellite-terrestrial network, where the trajectory and transmit power of UAVs were jointly optimized under interference temperature constraints. The authors of \cite{Wang2019} investigated the non-orthogonal multiple access (NOMA) strategy to integrate UAVs into the satellite
network. 
}
\par{
Despite of these achievements, there remain open challenges for CSUNs to efficiently support massive access out of the cellular coverage. On one hand, in a wide area, it is cost-ineffective to serve a massive number of IoT devices by deploying UAVs under conventional cellular architecture \cite{Lyu2018}, which motivates the design of cell-free on-demand coverage for CSUNs. On the other hand, to make the network be focused on scheduled IoT devices within a vast area, multi-domain resources, including subchannels, transmit power and hovering times, should be allocated jointly rather than in a separated or partially joint manner \cite{Bushnaq2019,Hua2019}, which needs the channel state information (CSI) of the whole system. However, the propagation condition is severe for wide-area IoT in practice, leading to much more complicated channel fading than the previously widely-used free-space path-loss model \cite{Hua2019}. This renders it necessary to study multi-domain resource allocation with imperfect CSI for CSUNs.
}
\subsection{Main Contributions}
\par{
In this paper, we consider a wide-area IoT-oriented CSUN consisting of a satellite and a swarm of UAVs. We jointly allocate the frequency-domain subchannels, transmit power and hovering time, aiming to establish a spectrum-efficient cell-free CSUN. Towards this end, a process-oriented optimization framework is proposed. Such framework takes the whole flight process of UAVs into account, which further derives multi-domain resource allocation schemes to improve network efficiency with guaranteed user fairness. Concretely, the main contributions are summarized as follows.
}
\begin{itemize}
\item We propose a process-oriented optimization framework by considering the whole UAV flight process for multi-domain resource allocation. The optimization is performed in a much larger time scale than channel coherent time, we thus use only the slowly-varying large-scale CSI, which can be predictively obtained according to the trajectory of UAV warm and position information of IoT devices. Besides, on-board energy constraints of UAV swarm and interference temperature constraints from UAVs to satellite users are also taken into account.
\item To improve network efficiency, we formulate a data transmission efficiency maximization problem under the process-oriented optimization framework. The original problem is decomposed into three subproblems, where subchannels, transmit power and hovering times are allocated by using the time-sharing relaxation method. Based on the solutions to these subproblems, the original problem is solved in an iterative way, leading to a low-complexity joint multi-domain resource allocation method. To further promote user fairness, i.e., offering services to all IoT devices as equally as possible, we formulate a minimum data transmission efficiency maximization problem under the process-oriented optimization framework. The problem is solved by similar decomposition and feasible region relaxation methods.
\item We evaluate the performance of the proposed algorithms by simulations. Particularly, the large-scale CSI is derived based on real geographical environment using channel models recommended by ITU-R \cite{ITU525,ITU676}. We observe an adaptive cell-free coverage pattern using the proposed multi-domain resource allocation algorithms. Moreover, although only the large-scale CSI is used for optimization, both network efficiency and user fairness can be improved significantly, due to the much enlarged time scale of optimization. 
\end{itemize}
\par{
The rest of this paper is organized as follows. We introduce the system model and the process-oriented optimization
framework in Section \uppercase\expandafter{\romannumeral2}. In Section \uppercase\expandafter{\romannumeral3}, the data transmission efficiency maximization problem is formulated and solved. We further discuss the minimum data transmission efficiency maximization problem and its solution in Section \uppercase\expandafter{\romannumeral4}.
Section \uppercase\expandafter{\romannumeral5} presents simulation results and discussions, and the conclusions are given in Section \uppercase\expandafter{\romannumeral6}.
}
\section{System Model}
\begin{figure*}[t]
\centering
\includegraphics[width=5.4in]{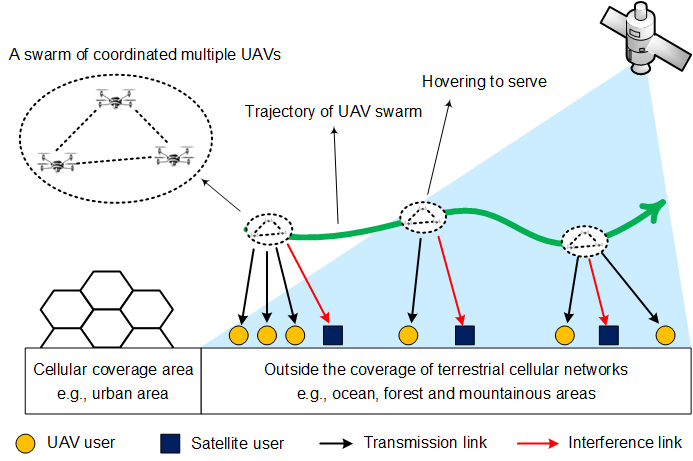}
\caption{ Illustration of a wide-area IoT-oriented CSUN outside the coverage of terrestrial cellular networks.}
\label{fig1}
\end{figure*}
 \par{
For future 6G networks, a massive number of IoT devices will be deployed globally. Thereby, we consider a wide-area IoT-oriented CSUN, which consists of a satellite and a swarm of coordinated $K$ single-antenna UAVs, serving $N_{s}$ satellite users and $N_{U}$ UAV users, as shown in Fig.~\ref{fig1}. We assume that each UAV user, as advanced IoT device, is equipped with $M$ antennas, and each satellite user, as general IoT device, is equipped with a single antenna. To support massive access with limited spectrum resources, UAV swarm and satellite share the same frequency band, which is divided into $G$ subchannels. UAV users are divided into $N$ groups \cite{Zhen2019}, and the $n$-th group of users will be served in the $n$-th time slot. Suppose that the $n$-th group has $U_n$ users, we have $\sum_{n = 1}^{N}U_n = N_{U}$. The UAV swarm works in a hover-to-serve mode, i.e., the UAVs transmit data when they are hovering above a group of users. After the accomplishment of services, they fly to the next user group. Such mode has advantages in high energy efficiency, high stability and small path loss for UAV communications \cite{Monwar2018, Bushnaq2019}, as it matches well the spatial sparsity of IoT devices in a wide area. We assume the hovering time of UAV swarm in the $n$-th time slot as $T_n$. Then, the hovering time constraints of UAV swarm are formulated by \cite{Liu2019}
\begin{equation}
 \sum_{n = 1}^{N} T_n \leq T_{total}  \label{totaltime} 
\end{equation}
\begin{equation}
 T_n \leq T_{max} \ \ \forall n \label{maxtime} 
\end{equation}
where (\ref{totaltime}) denotes the constraint of total hovering time and (\ref{maxtime}) shows the maximum available hovering time of UAVs in the $n$-th time slot.}
\par{ The received signal of the $u$-th UAV user in the $n$-th user group using the $g$-th subchannel can be expressed as
    \begin{equation}
        \mathbf{r}_{n, u, g} =  \mathbf{H}_{n,u,g} \mathbf{t}_{n,g} + \mathbf{q}_{n,u,g} \label{signalmodel1}
    \end{equation}
    where $n\in\{1,...,N\}$, $u\in\{1,...,U_n\}$, $g\in\{1,...,G\}$, $\mathbf{H}_{n,u,g} \in \mathbb{C}^{M \times K}$ denotes the channel matrix, $\mathbf{t}_{n,g} \in \mathbb{C}^{K}$ includes the transmitted symbols of UAV swarm and $\mathbf{q}_{n,u,g}\in \mathbb{C}^{M}$ denotes the additive white Gaussian noise following $\mathcal{CN}(\mathbf{0}_{M}, \sigma^{2}\mathbf{I}_{M})$, where $\mathbf{0}_{M} \in \mathbb{C}^{M}$ and $\mathbf{I}_{M} \in \mathbb{C}^{M \times M}$ are all-zero vector and identity matrix, respectively. Note that the leakage interference from the satellite has been ignored in (\ref{signalmodel1}), as it is relatively weak for advanced IoT devices. On the contrary, the leakage interference from the UAV swarm to satellite user is significant, which is
\begin{align}  
    \nonumber     \mathcal{I}_{n, i}  &  =  \sum_{u = 1}^{U_n} \sum_{g = 1}^{G}  x_{n,u,g} y_{n,i,g} \mathbf{h}_{n,i,g} \mathbf{E} \left\{ \mathbf{t}_{n, g} \mathbf{t}^{H}_{n, g} \right\} \mathbf{h}^{H}_{n,i,g}  \\
 & =  \sum_{u = 1}^{U_n} \sum_{g = 1}^{G}  x_{n,u,g} y_{n,i,g} \mathbf{h}_{n,i,g} \mathbf{P}_{n, g} \mathbf{h}^{H}_{n,i,g} \ \ \forall n,i \label{neinterference} 
\end{align}
where $n$ shows that the interference occurs in the $n$-th time slot, $i \in \{1,...,N_s\}$ are identifiers of satellite users, $x_{n, u, g} \in \{0, 1\}$ are indicator variables, $x_{n, u, g} = 1$ means that the $g$-th subchannel is used by the $u$-th UAV user of the $n$-th user group, $y_{n,i,g} \in \{0, 1\}$ are also indicator variables, $y_{n, i, g} = 1$ means that the $g$-th subchannel is used by the $i$-th satellite user in the $n$-th time slot, $\mathbf{h}_{n,i,g} \in \mathbb{C}^{1 \times K}$ denotes the channel vector of the interference link between the UAV swarm and the $i$-th satellite user in the $n$-th time slot using the $g$-th subchannel, $\mathbf{E} \left\{ \mathbf{t}_{n, g} \mathbf{t}^{H}_{n, g} \right\}$ is the correlation matrix of transmitted symbols. Generally, orthogonal symbols are transmitted by UAV swarm, so that $\mathbf{P}_{n, g} = \text{diag} \left\{p_{n, g, 1},...,p_{n, g, K} \right\}$ are diagonal matrices for $\forall n, g$ to represent the transmit power of UAV swarm in the $n$-th time slot using the $g$-th subchannel. 
}
\par{
  We consider a practical UAV channel model, including both line-of-sight (LOS) and non-line-of-sight (NLOS) elements, given by \cite{Chen2018,CXWang2018}
\begin{equation}
 \mathbf{H}_{n, u, g} =  \mathbf{S}_{n, u, g} \mathbf{L}_{n, u ,g}  \label{channelmodel1} 
\end{equation}
where $\mathbf{S}_{n, u, g}$ and $\mathbf{L}_{n, u, g}$ denote the small-scale fading and the slowly-varying large-scale fading, respectively. Particularly, the components of $\mathbf{S}_{n, u, g} \in \mathbb{C}^{M \times K}$ are independent and identically distributed (i.i.d.) standard complex Gaussian random variables, $\mathbf{L}_{n, u, g} = \text{diag} \left\{l_{n, u, g, 1},...,l_{n, u, g, K} \right\} \in \mathbb{R}^{K \times K}$, where $l^2_{n, u, g, k}$ represents the path loss between the $k$-th UAV and the $u$-th UAV user in the $n$-th time slot using the $g$-th subchannel. We assume that the interference link from UAV swarm to satellite user also follows the same channel model as
   \begin{equation}
   \mathbf{h}_{n,i,g} =  \mathbf{s}_{n, i, g} \widetilde{\mathbf{L}}_{n, i, g}  \label{channelmodel_sat} 
    \end{equation}
where $\mathbf{s}_{n,i,g}$ and $\widetilde{\mathbf{L}}_{n,i,g}$ denote the small-scale fading and the slowly-varying large-scale fading, respectively, $\mathbf{s}_{n,i,g} \in \mathbb{C}^{1 \times K}$ consists of i.i.d. standard complex Gaussian random variables, and $\widetilde{\mathbf{L}}_{n, i, g} = \text{diag} \left\{\tilde{l}_{n, i, g, 1},...,\tilde{l}_{n, i, g, K} \right\} \in \mathbb{R}^{K \times K}$ with $\tilde{l}^2_{n, i, g, k}$ representing the path loss between the $k$-th UAV and the $i$-th satellite user in the $n$-th time slot using the $g$-th subchannel. To be practical, we use the space-air channel models in Recommendation ITU-R P.525 and Recommendation ITU-R P.676 to derive $\mathbf{L}_{n, u, g}$ and $\widetilde{\mathbf{L}}_{n, i, g}$ based on real geographical information \cite{ITU525, ITU676}. 
}
\par{
We focus on radio resource allocation, thus assume an arbitrarily given trajectory of UAVs.\footnote{Resource allocation and trajectory planning of UAVs can be jointly optimized for CSUNs based on the results of this paper, which is an interesting future direction.} As the time scale of the whole UAV flight is much larger than the channel coherent time, it is impractical to acquire full CSI. We use the position-related large-scale CSI, i.e., $\mathbf{L}_{n, u, g}$ in (\ref{channelmodel1}) and $\widetilde{\mathbf{L}}_{n, i, g}$ in (\ref{channelmodel_sat}), for resource allocation, which can be predictively obtained according to trajectory and user locations \cite{Wangxx2019}. Using large-scale CSI, multi-domain resources can be allocated in an offline manner prior to UAV take-off, taking the whole flight process into account. This leads to a process-oriented optimization framework, under which the resource allocation is designed in a large time scale. Accordingly the leakage interference, network metrics and practical constraints should be derived in large-scale forms.
\par{Based on (\ref{neinterference}), we rewrite the leakage interference from UAV swarm to satellite user as
\begin{align} 
	\nonumber   \mathcal{I}^{e}_{n, i}  =  &  \mathbf{E}_{\mathbf{s}} \{  \mathcal{I}_{n, i} \} \\
    \nonumber    =   & \sum_{u = 1}^{U_n} \sum_{g = 1}^{G}  x_{n,u,g} y_{n,i,g} \\
\nonumber &   \mathbf{E}_{\mathbf{s}_{n,i,g}} \left\{ \mathbf{s}_{n,i,g} \widetilde{\mathbf{L}}_{n,i,g} \mathbf{P}_{n, g} \widetilde{\mathbf{L}}_{n,i,g} \mathbf{s}^{H}_{n,i,g} \right\} \\
  =  & \sum_{u = 1}^{U_n} \sum_{k = 1}^{K}  \sum_{g = 1}^{G} x_{n,u,g} y_{n,i,g} \tilde{l}_{n, i, g, k}^{2} p_{n, g, k} \ \ \forall n, i \label{interference} 
\end{align}
where $\mathbf{s} = \left\{\mathbf{s}_{n, i, g} \ \ \forall n,i,g \right\}$ is the set of small-scale channel parameters, $\mathbf{E}_{\mathbf{s}}$ represents the expectation with respect to small-scale parameters. Moreover, network efficiency and user fairness are important metrics to evaluate the performance of IoT-oriented CSUNs. From the network efficiency perspective, we focus on the overall data transmission efficiency, which is expressed as \cite{Feng2013}
 \begin{equation}
         \mathcal{D}_{e}(\mathbf{P}, \mathbf{T}, \mathbf{x}) = \mathbf{E}_{\mathbf{S}} \left\{\sum_{n = 1}^{N} \sum_{u = 1}^{U_n} \sum_{g = 1}^{G} x_{n, u, g} T_{n} R_{n, u, g} \right\} \label{De}
        \end{equation}
        where $\mathbf{P} = \{\mathbf{P}_{n, g} \ \ \forall n, g \}$ is the set of power matrices, $\mathbf{T} =(T_1,...,T_N)^T$, $\mathbf{x} = \{x_{n, u, g} \ \ \forall n,u,g\}$ is the set of indicator variables, $\mathbf{S} = \left\{\mathbf{S}_{n, u, g} \ \ \forall n,u,g \right\}$ denotes the set of small-scale channel parameters, $\mathbf{E}_{\mathbf{S}}$ represents the expectation with respect to small-scale parameters, and 
\begin{equation}
R_{n, u, g} = \text{log}_{2} \text{det} \left( \mathbf{I}_{M} + \frac{1}{\sigma^{2}} \mathbf{S}_{n, u, g} \mathbf{L}_{n, u, g} \mathbf{P}_{n, g} \mathbf{L}_{n, u, g} \mathbf{S}^{H}_{n, u, g} \right) \label{Rnug}
\end{equation}
denotes the downlink rate of the $u$-th UAV user in the $n$-th user group using the $g$-th subchannel. From the user fairness perspective, we consider the minimum data transmission efficiency of all users as
 \begin{equation}
         \mathcal{D}_{min}(\mathbf{P}, \mathbf{T}, \mathbf{x}) = \min _{n, u} \mathbf{E}_{\mathbf{S}} \left\{ \sum_{g = 1}^{G} x_{n, u, g} T_{n} R_{n, u, g} \right\}. \label{Dmin}
        \end{equation}
 For practical constraints, the on-board energy, transmit power and hovering time of the UAV flight process are regarded. Considering both propulsion energy and communication energy, we formulate the on-board energy constraints of UAV swarm as \cite{Liu2019}
\begin{equation}
 \frac{c}{\eta_{k}} \sum_{n = 1}^{N} \sum_{u = 1}^{U_n} \sum_{g = 1}^{G}  x_{n, u, g} p_{n, g, k} T_n +  E^{ind}_{k} + E^{prop}_{k} \leq E^{ob}_k \ \ \forall k \label{obenergy} 
\end{equation}
where $c$ is the power loss coefficient, $\eta_{k}$ denotes the efficiency of power amplifiers in radio frequency chains, $E^{ind}_{k}$ is the transmit-power-independent energy, including e.g., the energy consumed by cooling systems \cite{Joung2014}, $E^{prop}_{k}$ represents the propulsion energy which varies with the trajectory of UAVs. Given trajectories, both $E^{ind}_{k}$ and $E^{prop}_{k}$ will be fixed. Hence, we simplify the constraints of communication energy as
\begin{equation}
 \sum_{n = 1}^{N} \sum_{u = 1}^{U_n} \sum_{g = 1}^{G}  x_{n, u, g} p_{n, g, k} T_n \leq E^{com}_{k} \ \ \forall k  \label{comenergy}
\end{equation}
where $E^{com}_{k} = \frac{(E^{ob}_k - E^{ind}_{k} - E^{prop}_{k})\eta_{k}}{c}$. The maximum transmit power constraint of each UAV is also considered as
\begin{equation}
 \sum_{u = 1}^{U_n} \sum_{g = 1}^{G}  x_{n, u, g} p_{n, g, k} \leq p_{max} \ \ \forall n,k. \label{power} 
\end{equation} 
In the following Section \uppercase\expandafter{\romannumeral3} and Section \uppercase\expandafter{\romannumeral4}, we optimize the network efficiency and further promote the user fairness under this process-oriented optimization framework.
}
 \section{Process-oriented Data Transmission Efficiency Maximization}
We first formulate a data transmission efficiency maximization problem as
\begin{subequations} \label{pro1}
 \begin{align} 
	 	\max_{\mathbf{P}, \mathbf{T}, \mathbf{x}} &  \  \mathcal{D}_{e}(\mathbf{P}, \mathbf{T}, \mathbf{x}) \label{pro1a} \\
 s.t. \ \  & \sum_{u = 1}^{U_n} \sum_{k = 1}^{K}  \sum_{g = 1}^{G}  x_{n, u, g} y_{n, i, g} \tilde{l}_{n, i, g, k}^2 p_{n, g, k} \leq \epsilon_p \ \ \forall n,i  \label{pro1b} \\
& \sum_{n = 1}^{N} \sum_{u = 1}^{U_n} \sum_{g = 1}^{G}  x_{n, u, g} p_{n, g, k} T_n \leq E^{com}_{k} \ \ \forall k \label{pro1c} \\ 
& \sum_{u = 1}^{U_n} \sum_{g = 1}^{G}  x_{n, u, g} p_{n, g, k} \leq p_{max}  \ \ \forall n, k \label{pro1d} \\ 
& \sum_{n = 1}^{N} T_n \leq T_{total} \label{pro1e} \\ 
& T_n \leq T_{max} \ \ \forall n  \label{pro1f} \\ 
&  \sum_{u = 1}^{U_n} x_{n, u, g} \leq 1 \ \ \forall n,g  \label{pro1g} \\ 
 & x_{n, u, g} \in \{0, 1 \} \ p_{n, g, k} \geq 0 \ T_n \geq 0 \ \ \forall n,u,g,k \label{pro1h}
	\end{align}
\end{subequations}
where $\epsilon_p$ in (\ref{pro1b}) denotes the interference temperature threshold, (\ref{pro1c})--(\ref{pro1f}) are practical constraints as discussed in (\ref{totaltime})--(\ref{power}), and (\ref{pro1g}) means that one subchannel can only be used by one UAV user to avoid harmful interference. The problem in (\ref{pro1}) is a mixed-integer nonlinear programming (MINLP) problem, which is not convex and hard to be solved directly. In the following, we simplify (\ref{pro1}) and solve it in an iterative way.
\subsection{Problem Transformation}
    \par{
       First, we formulate a new objective function $\mathcal{D}_{a}(\mathbf{P},\mathbf{T}, \mathbf{w}, \mathbf{x})$ to closely approximate $\mathcal{D}_{e}(\mathbf{P}, \mathbf{T}, \mathbf{x})$ without expectation as follows.
       \begin{equation}
            \mathcal{D}_{a}(\mathbf{P},\mathbf{T},\mathbf{w}, \mathbf{x})  = \sum_{n = 1}^{N}  \sum_{u = 1}^{U_n}  \sum_{g = 1}^{G}  x_{n, u, g}T_n R_a(\mathbf{P}_{n, g}, w_{n, u ,g}) \label{Da}
        \end{equation}
where 
\begin{align}
\nonumber  R_a(\mathbf{P}_{n, g}, w_{n, u ,g}) = & \sum_{k = 1}^{K} \text{log}_{2} \left(1  + \frac{M l_{n, u, g, k}^{2} p_{n, g, k}}{w_{n, u, g} \sigma^{2}} \right)  \\
& +  M \left[ \text{log}_{2}w_{n, u, g} - \text{log}_{2}e (1 - w_{n, u, g}^{-1}) \right] \label{Rapw}
\end{align}
and $\mathbf{w} = \{w_{n, u, g}  \ \ \forall n,u,g \}$ is a set of slack variables which satisfies
        \begin{equation}
            w_{n, u, g} = 1 + \sum_{k = 1}^{K} \frac{l_{n,u,g,k}^{2} p_{n, g, k}}{\sigma^{2} + Ml_{n, u, g, k}^{2} p_{n, g, k} w_{n, u, g}^{-1}}. \label{slackw}
        \end{equation}
 The equation in (\ref{Rapw}) shows that the approximate rate is a sum of modified data rates and compensation terms, both of which are related to $\mathbf{w}$. The equation in (\ref{slackw}) indicates that $\mathbf{w}$ is an intractable implicit function of $\mathbf{P}$. The accuracy of this approximation technique has been discussed in \cite{Feng2013} in details. Thus, we recast (\ref{pro1}) as
        \begin{subequations} \label{pro2}
        \begin{align}
            \max_{\mathbf{P}, \mathbf{T}, \mathbf{x}} & \ \  \mathcal{D}_{a}(\mathbf{P}, \mathbf{T}, \mathbf{w},\mathbf{x}) \label{pro2a} \\
s.t. \ \  & \sum_{u = 1}^{U_n} \sum_{k = 1}^{K} \sum_{g = 1}^{G}   x_{n, u, g} y_{n, i, g} \tilde{l}_{n, i, g, k}^2 p_{n, g, k} \leq \epsilon_p \ \ \forall n,i \label{pro2b} \\
& \sum_{n = 1}^{N} \sum_{u = 1}^{U_n} \sum_{g = 1}^{G}  x_{n, u, g} p_{n, g, k} T_n \leq E^{com}_{k} \ \ \forall k \label{pro2c} \\ 
&\sum_{u = 1}^{U_n} \sum_{g = 1}^{G}  x_{n, u, g} p_{n, g, k} \leq p_{max} \ \ \forall n,k \label{pro2d} \\ 
& \sum_{n = 1}^{N} T_n \leq T_{total} \label{pro2g} \\ 
& T_n \leq T_{max} \ \ \forall n  \label{pro2h} \\ 
&  \sum_{u = 1}^{U_n} x_{n, u, g} \leq 1 \ \ \forall n,g \label{pro2e} \\ 
\nonumber &   w_{n, u, g}  \\
&= 1 + \sum_{k = 1}^{K} \frac{l_{n, u, g, k}^{2} p_{n, g, k}}{\sigma^{2} + Ml_{n, u, g, k}^{2} p_{n, g, k} w_{n, u, g}^{-1}}  \  \ \forall n,u,g \label{pro2i} \\
 & x_{n, u, g} \in \{0, 1 \} \ p_{n, g, k} \geq 0  \ T_n \geq 0 \ \ \forall n,u,g,k \label{pro2j}
        \end{align}
        \end{subequations}
        where (\ref{pro2i}) is introduced by the coupling between $\mathbf{P}$ and $\mathbf{w}$ as shown in (\ref{slackw}).}
        \subsection{Problem Decomposition}
        \par{
        The new problem in (\ref{pro2}) is not convex. To solve it, we decompose (\ref{pro2}) into three subproblems, following the block coordinate descent method \cite{Wu2018}. Denoting the iteration index as $r$, three subproblems are formulated as
        \begin{subequations} \label{subprox}
        \begin{align}
            \max_{\mathbf{x}^{r}} & \ \   \mathcal{D}_{a}(\mathbf{P}^{r - 1},\mathbf{T}^{r - 1}, \mathbf{w}^{r - 1}, \mathbf{x}^{r}) \label{subproxa} \\
  s.t. \ \ &  \sum_{u = 1}^{U_n} \sum_{k = 1}^{K} \sum_{g = 1}^{G}   x^{r}_{n, u, g} y_{n, i, g} \tilde{l}_{n, i, g, k}^2 p^{r - 1}_{n, g, k} \leq \epsilon_p  \ \ \forall n,i \label{subproxb} \\
& \sum_{n = 1}^{N} \sum_{u = 1}^{U_n} \sum_{g = 1}^{G}  x^{r}_{n, u, g} p^{r - 1}_{n, g, k} T^{r - 1}_n \leq E^{com}_{k} \ \ \forall k \label{subproxc} \\ 
&\sum_{u = 1}^{U_n} \sum_{g = 1}^{G}  x^{r}_{n, u, g} p^{r - 1}_{n, g, k} \leq p_{max} \ \ \forall n,k \label{subproxd} \\ 
&  \sum_{u = 1}^{U_n} x^{r}_{n, u, g} \leq 1 \ \ \forall n,g  \label{subproxe} \\ 
& x^{r}_{n, u, g} \in \{0, 1 \} \ \ \forall n,u,g  \label{subproxg}
        \end{align}
        \end{subequations}
        \begin{subequations} \label{subpro1}
        \begin{align}
            \max_{\mathbf{P}^{r}} & \ \   \mathcal{D}_{a}(\mathbf{P}^{r},\mathbf{T}^{r - 1}, \mathbf{w}^{r}, \mathbf{x}^{r}) \label{subpro1a} \\
  s.t. \ \  & \sum_{u = 1}^{U_n} \sum_{k = 1}^{K} \sum_{g = 1}^{G}   x^{r}_{n, u, g} y_{n, i, g} \tilde{l}_{n, i, g, k}^2 p^{r}_{n, g, k} \leq \epsilon_p \ \ \forall n,i \label{subpro1b} \\
& \sum_{n = 1}^{N} \sum_{u = 1}^{U_n} \sum_{g = 1}^{G}  x^{r}_{n, u, g} p^{r}_{n, g, k} T^{r - 1}_n \leq E^{com}_{k} \ \ \forall k \label{subpro1c} \\ 
 &\sum_{u = 1}^{U_n} \sum_{g = 1}^{G}  x^{r}_{n, u, g} p^{r}_{n, g, k} \leq p_{max} \ \ \forall n,k \label{subpro1d} \\ 
 \nonumber & w^{r}_{n, u, g} \\
&= 1 + \sum_{k = 1}^{K} \frac{l_{n,u,g,k}^{2} p^{r}_{n, g, k}}{\sigma^{2} + Ml_{n,u,g,k}^{2} p^{r}_{n, g, k} (w^{r}_{n, u, g})^{-1}} \ \ \forall n,u,g \label{subpro1e} \\
& p^r_{n, g, k} \geq 0 \ \ \forall n,g,k \label{subpro1f}
        \end{align}
        \end{subequations}
        \begin{subequations} \label{subpro2}
        \begin{align}
            \max_{\mathbf{T}^{r}} & \ \   \mathcal{D}_{a}(\mathbf{P}^{r},\mathbf{T}^{r}, \mathbf{w}^{r}, \mathbf{x}^{r}) \label{subpro2a} \\
         s.t. \   & \sum_{n = 1}^{N} \sum_{u = 1}^{U_n} \sum_{g = 1}^{G}  x^{r}_{n, u, g} p^{r}_{n,g,k} T^{r}_n \leq E^{com}_{k} \ \ \forall k \label{subpro2b} \\ 
& \sum_{n = 1}^{N} T^{r}_{n} \leq T_{total}  \label{subpro2c} \\
            & 0 \leq T^{r}_{n} \leq T_{max} \ \ \forall n. \label{subpro2d}
        \end{align}
        \end{subequations}
The subproblem in (\ref{subprox}) is an integer linear programming (ILP) problem, and its solution is referred to as the subchannel allocation scheme. The subproblem in (\ref{subpro1}) is non-convex, whose solution is referred to as the coordinated power allocation scheme. The subproblem in (\ref{subpro2}) is a linear programming problem, which can be directly solved using linear optimization tools \cite{Boyd2004Convex}. Its solution is referred to as the hovering time scheduling scheme. We solve (\ref{pro2}) iteratively via solving these subproblems in a turbo fashion, and focus on deriving the solutions to (\ref{subprox}) and (\ref{subpro1}). The methods will be described in Section \uppercase\expandafter{\romannumeral3}-C and Section \uppercase\expandafter{\romannumeral3}-D. 
}
\subsection{Subchannel Allocation}
\par{
We use the time-sharing relaxation technique \cite{Wei2019} to solve the subproblem in (\ref{subprox}). Concretely, $x^{r}_{n, u, g} \in \{0, 1 \}$ is relaxed to continuous $z^{r}_{n, u, g} \in [0, 1]$. Actually, $z^{r}_{n, u, g}$ can be regarded as the fraction of time that is used by the $u$-th UAV user in the $n$-th user group at the $g$-th subchannel. Then, we recast (\ref{subprox}) as 
        \begin{subequations} \label{subproz}
        \begin{align}
            \max_{\mathbf{z}^{r}} & \ \   \mathcal{D}_{a}(\mathbf{P}^{r - 1},\mathbf{T}^{r - 1}, \mathbf{w}^{r - 1}, \mathbf{z}^{r}) \label{subproza} \\
 s.t. \ \  & \sum_{u = 1}^{U_n} \sum_{k = 1}^{K} \sum_{g = 1}^{G}   z^{r}_{n, u, g} y_{n, i, g} \tilde{l}_{n, i, g, k}^2 p^{r - 1}_{n, g, k} \leq \epsilon_p \ \ \forall n,i \label{subprozb} \\
& \sum_{n = 1}^{N} \sum_{u = 1}^{U_n} \sum_{g = 1}^{G}  z^{r}_{n, u, g} p^{r - 1}_{n, g, k} T^{r - 1}_n \leq E^{com}_{k} \ \ \forall k \label{subprozc} \\ 
&\sum_{u = 1}^{U_n} \sum_{g = 1}^{G}  z^{r}_{n, u, g} p^{r - 1}_{n, g, k} \leq p_{max} \ \ \forall n,k \label{subprozd} \\ 
&  \sum_{u = 1}^{U_n} z^{r}_{n, u, g} \leq 1 \ \ \forall n, g  \label{subproze} \\ 
& 0 \leq z^{r}_{n, u, g} \leq 1 \ \ \forall n,u,g  \label{subprozg}
        \end{align}
        \end{subequations}
which is a linear programming problem that can be solved using linear optimization tools \cite{Boyd2004Convex}. Then, the key point is how to find $\mathbf{x}^{r}$ by using $\mathbf{z}^{r}$.
}
\par{
For this purpose, we formulate the Lagrangian dual function of (\ref{subproz}) as
\begin{align} \label{Lag}
\nonumber & L(\mathbf{z}^{r}, \bm{\lambda}, \bm{\mu}, \bm{\gamma}, \bm{\zeta}) = \mathcal{D}_{a}(\mathbf{P}^{r - 1}, \mathbf{T}^{r - 1}, \mathbf{w}^{r - 1}, \mathbf{z}^{r}) \\
\nonumber & + \sum_{n = 1}^{N} \sum_{i = 1}^{N_s} \lambda_{n, i} (\epsilon_p - \sum_{u = 1}^{U_n} \sum_{k = 1}^{K} \sum_{g = 1}^{G}   z^{r}_{n, u, g} y_{n, i, g} \tilde{l}_{n, i, g, k}^2 p^{r - 1}_{n, g, k}) \\
\nonumber & + \sum_{k = 1}^{K} \mu_{k} ( E^{com}_{k} - \sum_{n = 1}^{N} \sum_{u = 1}^{U_n} \sum_{g = 1}^{G}  z^{r}_{n, u, g} p^{r - 1}_{n, g, k} T^{r - 1}_n) \\
\nonumber & + \sum_{n = 1}^{N} \sum_{k = 1}^{K} \gamma_{n, k}(p_{max} - \sum_{u = 1}^{U_n} \sum_{g = 1}^{G}  z^{r}_{n, u, g} p^{r - 1}_{n, g, k}) \\
 & + \sum_{n = 1}^{N} \sum_{g = 1}^{G} \zeta_{n, g} (1 - \sum_{u = 1}^{U_n} z_{n,u,g}^{r}) 
\end{align}
where $\bm{\lambda}, \bm{\mu}, \bm{\gamma}, \bm{\zeta}$ are Lagrangian multipliers and the Lagrangian dual problem of (\ref{subproz}) is derived as
 \begin{subequations} \label{dualpro}
\begin{align}
\min_{\bm{\lambda}, \bm{\mu}, \bm{\gamma}, \bm{\zeta}} & \ f(\bm{\lambda}, \bm{\mu}, \bm{\gamma}, \bm{\zeta}) \\
 s.t. \ & \lambda_{n, i} \geq 0 \ \mu_{k} \geq 0 \ \gamma_{n, k} \geq 0 \ \zeta_{n, g} \geq 0 \ \ \forall n,i,k,g
\end{align}
 \end{subequations}
where
\begin{equation}\label{funcf}
f(\bm{\lambda}, \bm{\mu}, \bm{\gamma},\bm{\zeta}) = \sup_{\mathbf{z}^{r}} L(\mathbf{z}^{r}, \bm{\lambda}, \bm{\mu}, \bm{\gamma}, \bm{\zeta})
\end{equation}
is the least upper bound of (\ref{Lag}). Based on (\ref{Lag})--(\ref{funcf}), the desired $\mathbf{x}^{r}$ can be obtained in an iterative way. Denoting the iteration index as $t$, $\mathbf{x}^{t}$ is derived by
\begin{equation} \label{selectx}
x^{t}_{n, u^{*}, g} = 
\begin{cases}
1, & \text{if} \ u^{*} = \arg \max_{u} \{ V^{t}_{n, u, g} \ \ \forall n,g \} \\
0, & \text{else}
\end{cases} 
\end{equation}
where 
\begin{align} \label{V}
\nonumber &	V^{t}_{n, u, g} = \frac{\partial L(\mathbf{z}^{r}, \bm{\lambda}^{t - 1}, \bm{\mu}^{t - 1}, \bm{\gamma}^{t - 1},\bm{\zeta}^{t - 1})}{\partial z^{r}_{n, u, g}} + \zeta^{t - 1}_{n, g} \\
\nonumber & =T^{r - 1}_n R_a(\mathbf{P}^{r - 1}_{n, g}, w^{r - 1}_{n, u ,g}) - \sum_{i = 1}^{N_s} \sum_{k = 1}^{K}  \lambda^{t - 1}_{n, i} y_{n, i, g} \tilde{l}_{n, i, g, k}^2 p^{r - 1}_{n, g, k} \\
& - \sum_{k = 1}^{K} \mu^{t - 1}_{k} p^{r - 1}_{n, g, k} T^{r - 1}_n - \sum_{k = 1}^{K} \gamma^{t - 1}_{n, k} p^{r - 1}_{n, g, k}
\end{align}
 is formulated using the Karush-Kuhn-Tucker (KKT) conditions of (\ref{dualpro}) \cite{Boyd2004Convex}, which is a sum of overall data transmission efficiency and penalty terms.} Then, the Lagrangian multipliers are updated iteratively using the subgradient method by
\begin{equation} \label{lambda}
 \lambda^{t}_{n, i} = \left[ \lambda^{t - 1}_{n, i}  + \delta_{1}^{t} \frac{\partial L(\mathbf{x}^{t}, \bm{\lambda}, \bm{\mu}, \bm{\gamma}, \bm{\zeta})}{\partial \lambda_{n, i}} \right]^{+}
\end{equation}
\begin{equation} \label{mu}
\mu^{t}_{k} = \left[ \mu^{t - 1}_{k}  + \delta_{2}^{t} \frac{\partial L(\mathbf{x}^{t}, \bm{\lambda}, \bm{\mu}, \bm{\gamma}, \bm{\zeta})}{\partial \mu_{k}} \right]^{+}
\end{equation}
\begin{equation} \label{gamma}
\gamma^{t}_{n, k} = \left[ \gamma^{t - 1}_{n, k}  + \delta_{3}^{t}\frac{\partial L(\mathbf{x}^{t}, \bm{\lambda}, \bm{\mu}, \bm{\gamma}, \bm{\zeta})}{\partial \gamma_{n, k}} \right]^{+}
\end{equation}
where $\mathbf{x}^{t}$ has been substituted into the Lagrangian function in (\ref{lambda})--(\ref{gamma}), $[\cdot]^{+} = \max(\cdot, 0)$ and
\begin{align}
\nonumber \frac{\partial L(\mathbf{x}^{t}, \bm{\lambda}, \bm{\mu}, \bm{\gamma}, \bm{\zeta})}{\partial \lambda_{n, i}}=  & \epsilon_p \\
& - \sum_{u = 1}^{U_n} \sum_{k = 1}^{K} \sum_{g = 1}^{G}  x^{t}_{n, u, g} y_{n, i, g} \tilde{l}_{n, i, g, k}^2 p^{r - 1}_{n, g, k}
\end{align}
\begin{equation}
\frac{\partial L(\mathbf{x}^{t}, \bm{\lambda}, \bm{\mu}, \bm{\gamma}, \bm{\zeta})}{\partial \mu_{k}} = E^{com}_{k} - \sum_{n = 1}^{N} \sum_{u = 1}^{U_n} \sum_{g = 1}^{G}  x^{t}_{n, u, g} p^{r - 1}_{n, g, k} T^{r - 1}_n
\end{equation}
\begin{equation} \label{partialend}
\frac{\partial L(\mathbf{x}^{t}, \bm{\lambda}, \bm{\mu}, \bm{\gamma}, \bm{\zeta})}{\partial \gamma_{n, k}} = p_{max} - \sum_{u = 1}^{U_n} \sum_{g = 1}^{G}  x^{t}_{n, u, g} p^{r - 1}_{n, g, k}.
\end{equation}
Based on (\ref{selectx})--(\ref{partialend}), the subchannels can be allocated by Algorithm 1. The convergence of this algorithm can be guaranteed when the input parameters are appropriately designed \cite{Wei2019}. In Algorithm 1, $x^{t}_{n, u, g} = 1$ if and only if $R_a(\mathbf{P}_{n, g}, w_{n, u ,g})$ in (\ref{V}) is the largest one for $\forall u$ at the $t$-th step of iteration, showing that $\mathcal{D}_{a}(\mathbf{P},\mathbf{T}, \mathbf{w}, \mathbf{x})$ is maximized at every step of iteration. Hence, at least a locally optimal solution to (\ref{subprox}) can be obtained by Algorithm 1.
}
\addtolength{\topmargin}{0.05in}
 \begin{algorithm}[t] 
\caption{Algorithm to solve (\ref{subprox})}
\hspace*{0.02in} {\bf Input:}
$\left\{E^{com}_{k} \ \ \forall k \right\}$, $p_{max}$, $\epsilon_{p}$, $\mathbf{T}^{r - 1}$, $\mathbf{P}^{r - 1}$, $\delta^{1}_{1}$, $\delta^{1}_{2}$, $\delta^{1}_{3}$.
\begin{algorithmic}[1]
\State Initialization: $\mathbf{x}^{0} = \mathbf{0}$, $\bm{\lambda}^{0} = \mathbf{0}$, $\bm{\mu}^{0} = \mathbf{0}$,$\bm{\gamma}^{0} = \mathbf{0}$, $t = 1$;
\Repeat
\State Calculate $V^{t}_{n,u,g}$ using (\ref{V});
\State Update $\mathbf{x}^{t}$ using (\ref{selectx});
\State Update $\bm{\lambda}^{t}$ using (\ref{lambda}), where $\delta^{t}_{1} = \delta^{1}_{1} / t$;
\State Update $\bm{\mu}^{t}$ using (\ref{mu}), where $\delta^{t}_{2} = \delta^{1}_{2} / t$;
\State Update $\bm{\gamma}^{t}$ using (\ref{gamma}), where $\delta^{t}_{3} = \delta^{1}_{3} / t$;
\State $t = t + 1$;
\Until{$\mathbf{x}^{t}$ does not change;}
\end{algorithmic}
\hspace*{0.02in} {\bf Output:}
 $\mathbf{x}^{t}$. \\
\end{algorithm}
\subsection{Coordinated Power Allocation}
\par{
       In this section, we give the solution to (\ref{subpro1}). The objective function in (\ref{subpro1a}) is convex when both $\mathbf{P}^{r}$ and $\mathbf{w}^{r}$ satisfy (\ref{subpro1e}) \cite{Feng2013}. However, the coupling between $\mathbf{P}^{r}$ and $\mathbf{w}^{r}$ is too complicated as shown in (\ref{subpro1e}), so that it is hard to solve (\ref{subpro1}) directly with low computational complexity. To reduce the complexity, we relax (\ref{subpro1e}) and then solve (\ref{subpro1}) in an iterative way. Denoting the iteration index as $j$, (\ref{subpro1}) can be recast to
        \begin{subequations} \label{subpro1.1}
        \begin{align}
            \max_{\mathbf{P}^{j}} & \ \mathcal{D}_{a}(\mathbf{P}^{j}, \mathbf{T}^{r - 1}, \mathbf{w}^{j - 1}, \mathbf{x}^{r}) \label{subpro1.1a} \\
 s.t. \ \  & \sum_{u = 1}^{U_n} \sum_{k = 1}^{K} \sum_{g = 1}^{G}   x^{r}_{n, u, g} y_{n, i, g} \tilde{l}_{n, i, g, k}^2 p^{j}_{n, g, k} \leq \epsilon_p \ \ \forall n,i  \label{subpro1.1b} \\
& \sum_{n = 1}^{N} \sum_{u = 1}^{U_n} \sum_{g = 1}^{G}  x^{r}_{n, u, g} p^{j}_{n, g, k} T^{r - 1}_n \leq E^{com}_{k} \ \ \forall k \label{subpro1.1c} \\ 
 &\sum_{u = 1}^{U_n} \sum_{g = 1}^{G}  x^{r}_{n, u, g} p^{j}_{n, g, k} \leq p_{max} \ \ \forall n,k \label{subpro1.1d} \\ 
  & p^{j}_{n, g, k} \geq 0 \ \ \forall n,g,k  \label{subpro1.1e}
        \end{align}
        \end{subequations}
where $\mathbf{w}^{j - 1}$ is regarded as constant in (\ref{subpro1.1}). After $\mathbf{P}^{j}$ is obtained, $\mathbf{w}^{j}$ is updated by solving
\begin{equation} \label{wj}
 w^{j}_{n, u, g} = 1 + \sum_{k = 1}^{K} \frac{l_{n,u,g,k}^{2} p^{j}_{n, g, k}}{\sigma^{2} + Ml_{n,u,g,k}^{2} p^{j}_{n, g, k} (w^{j}_{n, u, g})^{-1}} \ \ \forall n,u,g.
\end{equation}
 Based on the solutions to (\ref{subpro1.1}) and (\ref{wj}), we can derive the solution to (\ref{subpro1}) using Algorithm 2. }
\par{
Then, we investigate the convergence of Algorithm 2. To this end, we first substitute $w^{r}_{n, u, g} = e^{v^{r}_{n,u,g}}$ into (\ref{subpro1}) to recast it as
\begin{subequations} \label{subpro1.2}
        \begin{align}
            \max_{\mathbf{P}^{r}} & \ \min_{\mathbf{v}^{r}} \  \mathcal{D}_{a}(\mathbf{P}^{r}, \mathbf{T}^{r - 1}, \mathbf{v}^{r}, \mathbf{x}^{r}) \label{subpro1.2a} \\
 s.t. \ \  & \sum_{u = 1}^{U_n} \sum_{k = 1}^{K} \sum_{g = 1}^{G}   x^{r}_{n, u, g} y_{n, i, g} \tilde{l}_{n, i, g, k}^2 p^{r}_{n, g, k} \leq \epsilon_p \ \ \forall n,i \label{subpro1.2b} \\
& \sum_{n = 1}^{N} \sum_{u = 1}^{U_n} \sum_{g = 1}^{G}  x^{r}_{n, u, g} p^{r}_{n, g, k} T^{r - 1}_n \leq E^{com}_{k} \ \ \forall k \label{subpro1.2c} \\ 
 &\sum_{u = 1}^{U_n} \sum_{g = 1}^{G}  x^{r}_{n, u, g} p^{r}_{n, g, k} \leq p_{max} \ \ \forall n,k \label{subpro1.2d} \\ 
 & p^{r}_{n, g, k} \geq 0 \ v^{r}_{n, u, g} \geq 0 \ \ \forall n,g,k,u \label{subpro1.2e}
        \end{align}
        \end{subequations}
where
 \begin{align}
\nonumber & \mathcal{D}_{a}(\mathbf{P}^{r},\mathbf{T}^{r - 1},\mathbf{v}^{r},\mathbf{x}^{r})  \\
& =  \sum_{n = 1}^{N}  \sum_{u = 1}^{U_n}  \sum_{g = 1}^{G}  x^{r}_{n, u, g}T^{r - 1}_n R_{a}(\mathbf{P}^{r}_{n, g}, v^{r}_{n, u, g}) \label{Da_v}
\end{align}
and 
\begin{align}
 \nonumber  R_{a}(\mathbf{P}^{r}_{n, g}, v^{r}_{n, u, g})   = & \sum_{k = 1}^{K} \textnormal{ log}_{2} \left(1  + \frac{M l_{n, u, g, k}^{2} p^r_{n, g, k}}{e^{v^{r}_{n, u, g}} \sigma^{2}} \right) \\
& +  M \textnormal{log}_{2}e (v^{r}_{n, u, g} - 1 + e^{-v^{r}_{n, u, g}}). \label{Ra_v}
\end{align}

The equivalence between (\ref{subpro1}) and (\ref{subpro1.2}) can be proved by [36, Theorem 1]. Then, (\ref{subpro1.2}) can be further decomposed into two subproblems as
 \begin{subequations} \label{subpro1.3}
        \begin{align}
            \max_{\mathbf{P}^{j}} \ & \mathcal{D}_{a}(\mathbf{P}^{j}, \mathbf{T}^{r - 1}, \mathbf{v}^{j-1}, \mathbf{x}^{r}) \label{subpro1.3a} \\
  s.t. \ \  & \sum_{u = 1}^{U_n} \sum_{k = 1}^{K} \sum_{g = 1}^{G}   x^{r}_{n, u, g} y_{n, i, g} \tilde{l}_{n, i, g, k}^2 p^{j}_{n, g, k} \leq \epsilon_p  \ \ \forall n,i \label{subpro1.3b} \\
& \sum_{n = 1}^{N} \sum_{u = 1}^{U_n} \sum_{g = 1}^{G}  x^{r}_{n, u, g} p^{j}_{n, g, k} T^{r - 1}_n \leq E^{com}_{k} \ \ \forall k \label{subpro1.3c} \\ 
 &\sum_{u = 1}^{U_n} \sum_{g = 1}^{G}  x^{r}_{n, u, g} p^{j}_{n, g, k} \leq p_{max} \ \ \forall n,k \label{subpro1.3d} \\ 
   & p^{j}_{n, g, k} \geq 0 \ \ \forall n,g,k \label{subpro1.3e}
        \end{align}
        \end{subequations}
\begin{subequations} \label{subpro1.4}
        \begin{align}
             \min_{\mathbf{v}^{j}} \  &  \mathcal{D}_{a}(\mathbf{P}^{j}, \mathbf{T}^{r - 1}, \mathbf{v}^{j}, \mathbf{x}^{r}) \label{subpro1.4a} \\
            s.t. \ & v^{j}_{n, u, g} \geq 0 \ \ \forall n,u,g. \label{subpro1.4b}
        \end{align}
        \end{subequations}
According to \cite{Liu2019}, (\ref{subpro1.3}) is equivalent to (\ref{subpro1.1}) and (\ref{subpro1.4}) is equivalent to (\ref{wj}). Thus, the solution to (\ref{subpro1.2}) is also found by Algorithm 2, which is equivalent to the solution to (\ref{subpro1}). We propose a theorem based on (\ref{subpro1.3}) and (\ref{subpro1.4}) to show that Algorithm 2 is guaranteed to converge.
\begin{theorem} \label{saddle}
Suppose
\begin{align}
\nonumber & L(\mathbf{P}^{r}, \mathbf{v}^r, \bm{\nu}, \bm{\xi}, \bm{\theta}) = \mathcal{D}_{a}(\mathbf{P}^{r}, \mathbf{T}^{r - 1}, \mathbf{v}^r, \mathbf{x}^{r}) \\
\nonumber & + \sum_{n = 1}^{N} \sum_{i = 1}^{N_s} \nu_{n, i} (\epsilon_p - \sum_{u = 1}^{U_n} \sum_{k = 1}^{K} \sum_{g = 1}^{G}   x^{r}_{n, u, g} y_{n, i, g} \tilde{l}_{n, i, g, k}^2 p^{r}_{n, g, k}) \\
\nonumber & + \sum_{k = 1}^{K} \xi_{k} ( E^{com}_{k} - \sum_{n = 1}^{N} \sum_{u = 1}^{U_n} \sum_{g = 1}^{G}  x^{r}_{n, u, g} p^{r}_{n, g, k} T^{r - 1}_n) \\
 & + \sum_{n = 1}^{N} \sum_{k = 1}^{K} \theta_{n, k}(p_{max} - \sum_{u = 1}^{U_n} \sum_{g = 1}^{G}  x^{r}_{n, u, g} p^{r}_{n, g, k})
\end{align}
is the Lagrangian dual function of (\ref{subpro1.3}) where $\bm{\nu}, \bm{\xi}, \bm{\theta}$ are Lagrangian multipliers, $\mathbf{P}^{r}$ and $\mathbf{v}^r$ satisfy (\ref{subpro1.2e}). Algorithm 2 will converge to the saddle point of $ L(\mathbf{P}^{r}, \mathbf{v}^r, \bm{\nu}, \bm{\xi}, \bm{\theta})$ where $\mathbf{P}^{r}$ and $\mathbf{v}^r$ are variables.
\end{theorem}
\begin{IEEEproof}
It is not difficult to observe that $\mathcal{D}_{a}(\mathbf{P}^{r}, \mathbf{T}^{r - 1}, \mathbf{v}^r, \mathbf{x}^{r})$ is concave with respect to $\mathbf{P}^{r}$ and convex with respect to $\mathbf{v}^r$. Thus, $L(\mathbf{P}^{r}, \mathbf{v}^r, \bm{\nu}, \bm{\xi}, \bm{\theta})$ is also concave with respect to $\mathbf{P}^{r}$ and convex with respect to $\mathbf{v}^r$, because (\ref{subpro1.2b})--(\ref{subpro1.2d}) are linear constraints with respect to $\mathbf{P}^{r}$. As a result, we can conclude that the solution to (\ref{subpro1.2}) is a saddle point of $L(\mathbf{P}^{r}, \mathbf{v}^r,\bm{\nu}, \bm{\xi}, \bm{\theta})$ where $\mathbf{P}^{r}$ and $\mathbf{v}^r$ are variables. For Algorithm 2, it follows the directions of subgradients at every step of iteration to find the solution to (\ref{subpro1.2}). Hence, Algorithm 2 will converge to this saddle point. 
\end{IEEEproof}
According to [36, Theorem 2], $\mathcal{D}_{a}(\mathbf{P}^{r}, \mathbf{T}^{r - 1}, \mathbf{v}^r, \mathbf{x}^{r})$ is non-decreasing along with iterations, so that at least a locally optimal solution is derived by Algorithm 2.
}
\addtolength{\topmargin}{0.05in}
 \begin{algorithm}[t] 
\caption{Algorithm to solve (\ref{subpro1})}
\hspace*{0.02in} {\bf Input:}
$\left\{E^{com}_{k} \ \ \forall k \right\}$, $p_{max}$, $\epsilon_{p}$, $\mathbf{T}^{r - 1}$, $\mathbf{x}^{r}$.
\begin{algorithmic}[1]
\State Initialization: $\epsilon_{0} = 1 \times 10^{-3}$, $j = 1$, $\mathbf{P}^{0} = \mathbf{0}$, $\mathbf{w}^{0} = \mathbf{1}$;
\State Solve (\ref{subpro1.1}), denoting the solution as $\mathbf{P}^{*}$, set $\mathbf{P}^{1} = \mathbf{P}^{*}$;
    \While{$|1 - \frac{ \mathcal{D}_{a}(\mathbf{P}^{j - 1},\mathbf{T}^{r - 1}, \mathbf{w}^{j - 1}, \mathbf{x}^{r})} { \mathcal{D}_{a}(\mathbf{P}^{j},\mathbf{T}^{r - 1}, \mathbf{w}^{j}, \mathbf{x}^{r})}| > \epsilon_{0}$ 
}
        \State Solve (\ref{wj}), denoting the solution as $\mathbf{w}^{*}$, set $\mathbf{w}^{j} = \mathbf{w}^{*}$;
        \State $j = j + 1$;
        \State Solve (\ref{subpro1.1}), denoting the solution as $\mathbf{P}^{*}$, set $\mathbf{P}^{j} = \mathbf{P}^{*}$;
    \EndWhile
\end{algorithmic}
\hspace*{0.02in} {\bf Output:}
 $\mathbf{P}^{j}$, $\mathbf{w}^{j}$. \\
\end{algorithm}
\par{Using the solutions to (\ref{subprox}), (\ref{subpro1}) and (\ref{subpro2}), we propose an iterative algorithm to solve (\ref{pro2}). The steps of this algorithm are summarized in Algorithm 3. }
\subsection{Convergence Analysis}
\par{
In this section, the convergence of Algorithm 3 is analyzed. Denoting $\mathbf{x}^{r - 1}$ as the solution to (\ref{subprox}), $\mathbf{P}^{r - 1}$ as the solution to (\ref{subpro1}) and $\mathbf{T}^{r - 1}$ as the solution to (\ref{subpro2}) at the $(r - 1)$-th step. At the $r$-th step of iteration, we first have $\mathbf{x}^{r}$ as the locally optimal solution after (\ref{subprox}) is solved. Hence, we have
        \begin{equation}
        \mathcal{D}_{a}(\mathbf{P}^{r - 1}, \mathbf{T}^{r - 1}, \mathbf{w}^{r - 1}, \mathbf{x}^{r}) \geq \mathcal{D}_{a}(\mathbf{P}^{r - 1}, \mathbf{T}^{r - 1}, \mathbf{w}^{r - 1}, \mathbf{x}^{r - 1}). \label{B.1}
        \end{equation}
         Then, after (\ref{subpro1}) is solved, we have $\mathbf{P}^{r}$ as the locally optimal solution, and $\textbf{w}^{r}$ can be accordingly calculated in (\ref{slackw}), which satisfies 
         \begin{equation}
        \mathcal{D}_{a}(\mathbf{P}^{r}, \mathbf{T}^{r - 1}, \mathbf{w}^{r}, \mathbf{x}^{r}) \geq \mathcal{D}_{a}(\mathbf{P}^{r - 1}, \mathbf{T}^{r - 1}, \mathbf{w}^{r - 1}, \mathbf{x}^{r}). \label{B.2}
        \end{equation}
Finally, the optimal hovering time $\mathbf{T}^{r}$ is acquired by solving (\ref{subpro2}), which satisfies
 \begin{equation}
        \mathcal{D}_{a}(\mathbf{P}^{r}, \mathbf{T}^{r}, \mathbf{w}^{r}, \mathbf{x}^{r}) \geq \mathcal{D}_{a}(\mathbf{P}^{r}, \mathbf{T}^{r - 1}, \mathbf{w}^{r}, \mathbf{x}^{r}). \label{B.3}
        \end{equation}
Thus, we have
        \begin{equation}
         \mathcal{D}_{a}(\mathbf{P}^{r}, \mathbf{T}^{r}, \mathbf{w}^{r}, \mathbf{x}^{r}) \geq \mathcal{D}_{a}(\mathbf{P}^{r - 1}, \mathbf{T}^{r - 1}, \mathbf{w}^{r - 1}, \mathbf{x}^{r - 1}) \label{B.4}
        \end{equation}
        showing that the objective function of (\ref{pro2}) keeps increasing at every step of the iteration, and it is upper bounded by the given resources. As a result, the convergence of Algorithm 3 is guaranteed, and at least a locally optimal solution can be derived using this algorithm. 
}
\begin{remark}
\textnormal{The subchannel allocation method in Section \uppercase\expandafter{\romannumeral3}-C implies that the UAV users in better channel environments have more chance to be served. Using this strategy, although the overall data transmission efficiency can be improved, the fairness among UAV users is not guaranteed. For example, if a UAV user stays in bad channel environment for a long time, it can hardly be served by UAVs, which shows the lack of user fairness. Such phenomenon inspires us to consider a new network metric for user fairness. 
}
\end{remark}
 \addtolength{\topmargin}{0.05in}
 \begin{algorithm}[t]
\caption{Data transmission efficiency maximization algorithm}
\hspace*{0.02in} {\bf Input:}
$T_{total}$, $T_{max}$, $\left\{ E^{com}_{k} \ \ \forall k \right\}$, $p_{max}$, $\epsilon_{p}$.
\begin{algorithmic}[1]
\State Initialization: $\epsilon_{0} = 1 \times 10^{-2}$, $r = 1$, $\mathbf{T}^{0} = (T_{total} / N) \mathbf{1}$, $\mathbf{P}^{0} = \mathbf{0}$;
\State Solve (\ref{subprox}), denoting the solution as $\mathbf{x}^{*}$, set $\mathbf{x}^{1} = \mathbf{x}^{*}$;
\State Solve (\ref{subpro1}), denoting the solution as $\mathbf{P}^{*}$, set $\mathbf{P}^{1} = \mathbf{P}^{*}$;
\State Solve (\ref{subpro2}), denoting the solution as $\mathbf{T}^{*}$, set $\mathbf{T}^{1} = \mathbf{T}^{*}$;
    \While{$|1 - \frac { \mathcal{D}_{a}(\mathbf{P}^{r - 1},\mathbf{T}^{r - 1}, \mathbf{w}^{r - 1}, \mathbf{x}^{r - 1})} {\mathcal{D}_{a}(\mathbf{P}^{r},\mathbf{T}^{r}, \mathbf{w}^{r}, \mathbf{x}^{r})}| > \epsilon_{0}$}
		 \State $r = r + 1$;
        \State Solve (\ref{subprox}), denoting the solution as $\mathbf{x}^{*}$, set $\mathbf{x}^{r} = \mathbf{x}^{*}$;
\State Solve (\ref{subpro1}), denoting the solution as $\mathbf{P}^{*}$, set $\mathbf{P}^{r} = \mathbf{P}^{*}$;
\State Solve (\ref{subpro2}), denoting the solution as $\mathbf{T}^{*}$, set $\mathbf{T}^{r} = \mathbf{T}^{*}$;
    \EndWhile
\end{algorithmic}
\hspace*{0.02in} {\bf Output:}
 $\mathbf{x}^{r}$, $\mathbf{P}^{r}$, $\mathbf{T}^{r}$. \\
\end{algorithm}

\section{Process-oriented Minimum Data Transmission Efficiency Maximization}

\subsection{Problem Formulation and Decomposition}
\par{
To improve user fairness, we use (\ref{Dmin}) as the objective function. Further by using the technique in Section \uppercase\expandafter{\romannumeral3}-A, the approximate form of (\ref{Dmin}) can be derived, and the minimum data transmission efficiency maximization problem is accordingly formulated as
\begin{subequations} \label{pro3}
 \begin{align} 
	 	\max_{\mathbf{P}, \mathbf{T}, \mathbf{x}} &  \ \min_{n, u} \  \sum_{g = 1}^{G} x_{n, u, g} T_{n}  R_{a}(\mathbf{P}_{n, g}, w_{n, u, g}) \label{pro3a}  \\ 
  s.t. \ \  & \sum_{u = 1}^{U_n} \sum_{k = 1}^{K} \sum_{g = 1}^{G}   x_{n, u, g} y_{n, i, g} \tilde{l}_{n, i, g, k}^2 p_{n, g, k} \leq \epsilon_p \ \ \forall n,i \label{pro3b} \\
& \sum_{n = 1}^{N} \sum_{u = 1}^{U_n} \sum_{g = 1}^{G}  x_{n, u, g} p_{n, g, k} T_n \leq E^{com}_{k} \ \ \forall k \label{pro3c} \\ 
&\sum_{u = 1}^{U_n} \sum_{g = 1}^{G}  x_{n, u, g} p_{n, g, k} \leq p_{max} \ \ \forall n, k \label{pro3d} \\ 
&  \sum_{u = 1}^{U_n} x_{n, u, g} \leq 1 \ \ \forall n,g  \label{pro3e} \\ 
& \sum_{n = 1}^{N} T_n \leq T_{total} \label{pro3f} \\ 
& T_n \leq T_{max} \ \ \forall n  \label{pro3g} \\ 
 \nonumber & w_{n, u, g}  \\
&= 1 + \sum_{k = 1}^{K} \frac{l_{n, u, g, k}^{2} p_{n, g, k}}{\sigma^{2} + Ml_{n, u, g, k}^{2} p_{n, g, k} w_{n, u, g}^{-1}} \ \ \forall n, u, g \label{pro3h} \\
 & x_{n, u, g} \in \{0, 1 \} \ p_{n, g, k} \geq 0 \ T_n \geq 0 \ \ \forall n,u,g,k \label{pro3i}
	\end{align}
\end{subequations}
where $R_{a}(\mathbf{P}_{n, g}, w_{n, u, g})$ has been defined in (\ref{Rapw}), and (\ref{pro3b})--(\ref{pro3i}) are the same as the constraints in (\ref{pro1}). The problem in (\ref{pro3}) is a max-min MINLP problem, which is not convex and hard to be solved directly. Then, we decompose it into three subproblems following the block coordinate descent method, similar to Section \uppercase\expandafter{\romannumeral3}-B. Denoting the iteration index as $r$, the subproblems are formulated as
        \begin{subequations} \label{subpro3-x}
        \begin{align}
            \max_{\mathbf{x}^{r}} & \ \min_{n, u} \  \sum_{g = 1}^{G} x^{r}_{n, u, g} T^{r - 1}_{n}  R_{a}(\mathbf{P}^{r - 1}_{n, g}, w^{r - 1}_{n, u, g}) \label{subpro3-xa} \\
  s.t. \ \  & \sum_{u = 1}^{U_n} \sum_{k = 1}^{K} \sum_{g = 1}^{G}   x^{r}_{n, u, g} y_{n, i, g} \tilde{l}_{n, i, g, k}^2 p^{r - 1}_{n, g, k} \leq \epsilon_p  \ \ \forall n,i \label{subpro3-xb} \\
& \sum_{n = 1}^{N} \sum_{u = 1}^{U_n} \sum_{g = 1}^{G}  x^{r}_{n, u, g} p^{r - 1}_{n, g, k} T^{r - 1}_n \leq E^{com}_{k} \ \ \forall k \label{subpro3-xc} \\ 
&\sum_{u = 1}^{U_n} \sum_{g = 1}^{G}  x^{r}_{n, u, g} p^{r - 1}_{n, g, k} \leq p_{max} \ \ \forall n,k \label{subpro3-xd} \\ 
&  \sum_{u = 1}^{U_n} x^{r}_{n, u, g} \leq 1 \ \ \forall n,g  \label{subpro3-xe} \\ 
& x^{r}_{n, u, g} \in \{0, 1 \} \ \ \forall n,u,g  \label{subpro3-xf}
        \end{align}
        \end{subequations}
        \begin{subequations} \label{subpro3-1}
        \begin{align}
            \max_{\mathbf{P}^{r}} & \   \min_{n, u} \  \sum_{g = 1}^{G} x^{r}_{n, u, g} T^{r - 1}_{n}  R_{a}(\mathbf{P}^{r}_{n, g}, w^{r}_{n, u, g}) \label{subpro3-1a} \\
 s.t. \ \  & \sum_{u = 1}^{U_n} \sum_{k = 1}^{K} \sum_{g = 1}^{G}   x^{r}_{n, u, g} y_{n, i, g} \tilde{l}_{n, i, g, k}^2 p^{r}_{n, g, k} \leq \epsilon_p  \ \ \forall n,i \label{subpro3-1b} \\
& \sum_{n = 1}^{N} \sum_{u = 1}^{U_n} \sum_{g = 1}^{G}  x^{r}_{n, u, g} p^{r}_{n, g, k} T^{r - 1}_n \leq E^{com}_{k} \ \ \forall k \label{subpro3-1c} \\ 
 &\sum_{u = 1}^{U_n} \sum_{g = 1}^{G}  x^{r}_{n, u, g} p^{r}_{n, g, k} \leq p_{max} \ \ \forall n,k \label{subpro3-1d} \\ 
 \nonumber &  w^{r}_{n, u, g}  \\
&=1 + \sum_{k = 1}^{K} \frac{l_{n,u,g,k}^{2} p^{r}_{n, g, k}}{\sigma^{2} + Ml_{n,u,g,k}^{2} p^{r}_{n, g, k} (w^{r}_{n, u, g})^{-1}}  \ \ \forall n,u,g \label{subpro3-1e} \\
& p^r_{n, g, k} \geq 0 \ \ \forall n,g,k \label{subpro3-1f}
        \end{align}
        \end{subequations}
        \begin{subequations} \label{subpro3-2}
        \begin{align}
            \max_{\mathbf{T}^{r}} & \ \min_{n, u} \  \sum_{g = 1}^{G} x^{r}_{n, u, g} T^{r}_{n}  R_{a}(\mathbf{P}^{r}_{n, g}, w^{r}_{n, u, g}) \label{subpro3-2a} \\
         s.t. \   & \sum_{n = 1}^{N} \sum_{u = 1}^{U_n} \sum_{g = 1}^{G}  x^{r}_{n, u, g} p^{r}_{n, g, k} T^{r}_n \leq E^{com}_{k} \ \ \forall k \label{subpro3-2b} \\ 
& \sum_{n = 1}^{N} T^{r}_{n} \leq T_{total}  \label{subpro3-2c} \\
            & 0 \leq T^{r}_{n} \leq T_{max} \ \ \forall n. \label{subpro3-2d}
        \end{align}
        \end{subequations}
For three subproblems, (\ref{subpro3-x}) is a max-min ILP problem and (\ref{subpro3-1}) is non-convex, both of which are hard to be solved directly, while (\ref{subpro3-2}) is a linear max-min optimization problem, which can be directly solved using conventional max-min optimization tools \cite{Murray1980}. Hence, we focus on giving the solutions to (\ref{subpro3-x}) and (\ref{subpro3-1}).
}
\subsection{Max-min Subchannel Allocation}
\par{
To solve (\ref{subpro3-x}), we define a slack variable $\tau$, which satisfies
\begin{equation}
\tau = \min_{n, u} \  \sum_{g = 1}^{G} x^{r}_{n, u, g} T^{r - 1}_{n}  R_{a}(\mathbf{P}^{r - 1}_{n, g}, w^{r - 1}_{n, u, g}). \label{tau}
\end{equation}
Then, (\ref{subpro3-x}) can be equivalently transformed to
\begin{subequations} \label{subpro3-x1}
        \begin{align}
            \max_{\mathbf{x}^{r}, \tau} & \ \tau \label{subpro3-x1a} \\
          s.t. \ \ & \sum_{g = 1}^{G} x^{r}_{n, u, g} T^{r - 1}_{n}  R_{a}(\mathbf{P}^{r - 1}_{n, g},w^{r - 1}_{n, u, g}) \geq \tau \ \ \forall n,u \label{subpro3-x1b} \\
  & \sum_{u = 1}^{U_n} \sum_{k = 1}^{K} \sum_{g = 1}^{G}   x^{r}_{n, u, g} y_{n, i, g} \tilde{l}_{n, i, g, k}^2 p^{r - 1}_{n, g, k} \leq \epsilon_p  \ \ \forall n,i \label{subpro3-x1c} \\
& \sum_{n = 1}^{N} \sum_{u = 1}^{U_n} \sum_{g = 1}^{G}  x^{r}_{n, u, g} p^{r - 1}_{n, g, k} T^{r - 1}_n \leq E^{com}_{k} \ \ \forall k \label{subpro3-x1d} \\ 
&\sum_{u = 1}^{U_n} \sum_{g = 1}^{G}  x^{r}_{n, u, g} p^{r - 1}_{n, g, k} \leq p_{max} \ \ \forall n,k \label{subpro3-x1e} \\ 
&  \sum_{u = 1}^{U_n} x^{r}_{n, u, g} \leq 1 \ \ \forall n,g  \label{subpro3-x1f} \\ 
& x^{r}_{n, u, g} \in \{0, 1 \} \ \ \forall n,u,g.  \label{subpro3-x1g}
        \end{align}
        \end{subequations}
We can observe that (\ref{subpro3-x1}) is a mixed-ILP (MILP) problem and hard to be solved directly. Thus, we propose a theorem to simplify it.
\begin{theorem}
	The optimal solution to (\ref{subpro3-x1}) will not change after (\ref{subpro3-x1c})--(\ref{subpro3-x1e}) are relaxed.
\end{theorem}
\begin{IEEEproof}
See Appendix A.
\end{IEEEproof}
Theorem 2 states that (\ref{subpro3-x1b}) has the highest priority compared with other constraints, as it is a transformed form of the original objective function. According to Theorem 2, we can recast (\ref{subpro3-x1}) to
\begin{subequations} \label{subpro3-x2}
        \begin{align}
            \max_{\mathbf{x}^{r}, \tau} & \ \tau \label{subpro3-x2a} \\
          s.t. \ \ & \sum_{g = 1}^{G} x^{r}_{n, u, g} T^{r - 1}_{n}  R_{a}(\mathbf{P}^{r - 1}_{n, g},w^{r - 1}_{n, u, g}) \geq \tau \ \ \forall n,u \label{subpro3-x2b} \\
&  \sum_{u = 1}^{U_n} x^{r}_{n, u, g} \leq 1 \ \ \forall n,g  \label{subpro3-x2c} \\ 
& x^{r}_{n, u, g} \in \{0, 1 \} \ \ \forall n,u,g.  \label{subpro3-x2d}
        \end{align}
        \end{subequations}

To solve (\ref{subpro3-x2}), we have a property to show the condition that the solution to (\ref{subpro3-x2}) must satisfy.
\begin{property}
	Suppose $\mathbf{x}^{r*}$ as a non-trivial solution to (\ref{subpro3-x2}), then $\mathbf{x}^{r*}$ must satisfy
\begin{equation}
\sum_{g = 1}^{G} x_{n,u,g}^{r} \geq 1 \ \ \forall n,u. \label{property1}
\end{equation}
Otherwise, if (\ref{property1}) is not satisfied, (\ref{subpro3-x2}) only has a trivial solution, where the maximum value of $\tau$ is $0$.
\end{property}
\begin{IEEEproof}
If (\ref{property1}) is satisfied, the conclusion of this property is naturally given. Then, if (\ref{property1}) is not satisfied, as the variables in $\mathbf{x}^{r}$ are either $0$ or $1$, there exists $n^* \in \{1,...,N\}$ and $u^* \in \{1,...,U_n^{*}\}$ which satisfy
\begin{equation}
\sum_{g = 1}^{G} x_{n^*,u^*,g}^{r} = 0
\end{equation}
which means $x_{n^*,u^*,g}^{r}$ is $0$ for all $g \in \{1,...,G\}$. Substituting $x_{n^*,u^*,g}^{r}$ into (\ref{subpro3-x2}), we can find that the maximum value of $\tau$ is $0$. Hence, the conclusion of Property 1 is given.
\end{IEEEproof}
Property 1 shows that each user must have at least one subchannel to use, indicating that the user in worst condition has the highest priority to be served. Using Property 1, a solution to (\ref{subpro3-x}) can be derived based on (\ref{subpro3-x2}) in a greedy manner, which is summarized in Algorithm~4. At every step of Algorithm 4, the minimum data transmission efficiency is improved by allocating the subchannels to the user in worst condition as much as possible. Hence, Algorithm~4 can converge to the locally optimal solution to (\ref{subpro3-x}).
}
\begin{remark}
\textnormal{Note that we have proposed two different methods to solve the subchannel allocation subproblem in Section \uppercase\expandafter{\romannumeral3}-C and the max-min subchannel allocation subproblem in Section \uppercase\expandafter{\romannumeral4}-B. The reason is that the algorithms are designed to accommodate the objective functions of different problems to achieve better performance. The subchannel allocation algorithm in Section \uppercase\expandafter{\romannumeral3}-C can maximize the overall data transmission efficiency, while the method in Section \uppercase\expandafter{\romannumeral4}-B can improve the minimum data transmission efficiency. 
}
\end{remark}
\addtolength{\topmargin}{0.05in}
 \begin{algorithm}[t]
\caption{Max-min subchannel allocation algorithm}
\hspace*{0.02in} {\bf Input:}
$\mathbf{T}^{r - 1}$, $\mathbf{P}^{r - 1}$, $\mathbf{w}^{r - 1}$.
\begin{algorithmic}[1]
\State Initialization: $\epsilon_{0} = 1 \times 10^{-3}$, $j = 1$, $\tau^{0} = 0$;
\State Initialize $\mathbf{x}^{1}$ according to \cite{Shen2005};
\State Define $V^{j}_{n, u} = \sum_{g = 1}^{G} x^{j}_{n, u, g} T^{r - 1}_{n} R_{a}(\mathbf{P}^{r - 1}_{n, g},w^{r - 1}_{n, u, g})$;
\State Set $\tau^{j} = \min_{n, u} V^{j}_{n, u}$;
\While{$|1 - \frac{\tau^{j - 1}}{ \tau^{j}}| > \epsilon_{0}$}
		\For{n = 1:N}
				\State Find the minimum value of $V^{j}_{n, u}$, denoting the index as $u^{*}$;
			 \State Find the maximum value of $V^{j}_{n, u}$ that satisfies the condition $\sum_{g = 1}^{G}  x^{j}_{n, u, g} > 1$, denoting the index as $u^{**}$;
			\State Define the index set as $I = \{g | x^{j}_{n, u^{**}, g} = 1\}$;
			\State Find $g^* = \text{arg} \min_{g \in I} T^{r - 1}_{n}  R_{a}(\mathbf{P}^{r - 1}_{n, g},w^{r - 1}_{n, u^{**}, g})$;
			\If{$V^{j}_{n, u^{*}} + T^{r - 1}_{n}  R_{a}(\mathbf{P}^{r - 1}_{n, g^{*}},w^{r - 1}_{n, u^{*}, g^{*}}) \leq  V^{j}_{n, u^{**}} - T^{r - 1}_{n}  R_{a}(\mathbf{P}^{r - 1}_{n, g^{*}},w^{r - 1}_{n, u^{**}, g^{*}})$}
				\State Set $x^{j}_{n, u^{*}, g^*} = 1$;
				\State Set $x^{j}_{n, u^{**}, g^*} = 0$;
			\EndIf		 
      \EndFor
		\State $j = j + 1$;
		\State Set $\tau^{j} = \min_{n, u} V^{j}_{n, u}$;
\EndWhile
\end{algorithmic}
\hspace*{0.02in} {\bf Output:}
 $\mathbf{x}^{j}$. \\
\end{algorithm}
\subsection{Max-min Power Allocation}
\par{To solve (\ref{subpro3-1}), we define the slack variable $\tau$ similar to (\ref{tau}),  then (\ref{subpro3-1}) is rewritten as
        \begin{subequations} \label{subpro3-1-1}
        \begin{align}
            \max_{\mathbf{P}^{r}, \tau} & \ \tau  \label{subpro3-1-1a}\\
          s.t. \ \ & \sum_{g = 1}^{G} x^{r}_{n, u, g} T^{r - 1}_{n}  R_{a}(\mathbf{P}^{r}_{n, g}, w^{r}_{n, u, g}) \geq \tau \ \ \forall n,u \label{subpro3-1-1b} \\
 & \sum_{u = 1}^{U_n} \sum_{k = 1}^{K} \sum_{g = 1}^{G}   x^{r}_{n, u, g} y_{n, i, g} \tilde{l}_{n, i, g, k}^2 p^{r}_{n, g, k} \leq \epsilon_p \ \ \forall n,i \label{subpro3-1-1c} \\
& \sum_{n = 1}^{N} \sum_{u = 1}^{U_n} \sum_{g = 1}^{G}  x^{r}_{n, u, g} p^{r}_{n, g, k} T^{r - 1}_n \leq E^{com}_{k} \ \ \forall k \label{subpro3-1-1d} \\ 
 &\sum_{u = 1}^{U_n} \sum_{g = 1}^{G}  x^{r}_{n, u, g} p^{r}_{n, g, k} \leq p_{max} \ \ \forall n,k \label{subpro3-1-1e} \\ 
\nonumber  &  w^{r}_{n, u, g}  \\
& = 1 + \sum_{k = 1}^{K} \frac{l_{n,u,g,k}^{2} p^{r}_{n, g, k}}{\sigma^{2} + Ml_{n,u,g,k}^{2} p^{r}_{n, g, k} (w^{r}_{n, u, g})^{-1}} \ \ \forall n,u,g \label{subpro3-1-1f} \\
& p^r_{n, g, k} \geq 0 \ \ \forall n,g,k. \label{subpro3-1-1g}
        \end{align}
        \end{subequations}
Note that (\ref{subpro3-1-1}) is non-convex, due to the coupling between $\mathbf{P}$ and $\mathbf{w}$ in (\ref{subpro3-1-1b}) and (\ref{subpro3-1-1f}). To handle this problem, we propose a theorem to recast (\ref{subpro3-1-1b}) and (\ref{subpro3-1-1f}).
\begin{theorem}
The constraints in (\ref{subpro3-1-1b}) and (\ref{subpro3-1-1f}) can be equivalently transformed to
\begin{align} 
 &\sum_{g = 1}^{G} x^{r}_{n, u, g} T^{r - 1}_{n}  R_{a}(\mathbf{P}^{r}_{n, g}, v^{r}_{n, u, g}) \geq \tau \  \ \forall n,g  \label{Ra_vc1} \\
& v^{r}_{n, u, g} \geq 0 \ \ \forall n, u, g \label{Ra_vc2}
\end{align}
where $v^{r}_{n,u,g} = \textnormal{log}( w^{r}_{n,u,g})$, $R_{a}(\mathbf{P}^{r}_{n, g}, v^{r}_{n, u, g})$ has been defined in (\ref{Ra_v}).
\end{theorem}
\begin{IEEEproof}
\textnormal{See Appendix B.}
\end{IEEEproof}
\par{ Theorem 3 shows that (\ref{subpro3-1-1b}) and (\ref{subpro3-1-1f}) can be equivalently replaced by (\ref{Ra_vc1}) and (\ref{Ra_vc2}), indicating that $\mathbf{P}^{r}$ and $\mathbf{v}^{r}$ can be decoupled in (\ref{subpro3-1-1}). Consequently, successive convex optimization method can be used to solve (\ref{subpro3-1-1}), but the computational overhead is too large \cite{Sun2017}. Hence, we solve (\ref{subpro3-1-1}) in an iterative way with low complexity. Denoting the iteration index as $j$, (\ref{subpro3-1-1}) is recast to
  \begin{subequations} \label{subpro3-1-2}
        \begin{align}
            \max_{\mathbf{P}^{j}, \tau^{j}} & \ \tau^{j}  \label{subpro3-1-2a}\\
          s.t. \ \ & \sum_{g = 1}^{G} x^{r}_{n, u, g} T^{r - 1}_{n}  R_{a}(\mathbf{P}^{j}_{n, g}, v^{j - 1}_{n, u, g}) \geq \tau^{j} \ \ \forall n,u \label{subpro3-1-2b} \\
 & \sum_{u = 1}^{U_n} \sum_{k = 1}^{K} \sum_{g = 1}^{G}   x^{r}_{n, u, g} y_{n, i, g} \tilde{l}_{n, i, g, k}^2 p^{j}_{n, g, k} \leq \epsilon_p \ \ \forall n,i \label{subpro3-1-2c} \\
& \sum_{n = 1}^{N} \sum_{u = 1}^{U_n} \sum_{g = 1}^{G}  x^{r}_{n, u, g} p^{j}_{n, g, k} T^{r - 1}_n \leq E^{com}_{k} \ \ \forall k \label{subpro3-1-2d} \\ 
 &\sum_{u = 1}^{U_n} \sum_{g = 1}^{G}  x^{r}_{n, u, g} p^{j}_{n, g, k} \leq p_{max} \ \ \forall n,k \label{subpro3-1-2e} \\ 
& p^j_{n, g, k} \geq 0 \ \ \forall n,g,k \label{subpro3-1-2f}
        \end{align}
        \end{subequations}
where $\mathbf{v}^{j - 1}$ is regarded as constant. Then, $\mathbf{v}^{j}$ is derived by solving the equations as follows.
\begin{equation} \label{vp}
e^{v^{j}_{n,u,g}} = 1 + \sum_{k = 1}^{K} \frac{l_{n,u,g,k}^{2}p_{n,g,k}^{j}}{\sigma^{2} + M l_{n,u,g,k}^{2}p_{n,g,k}^{j}e^{-v^{j}_{n,u,g}}} \ \ \forall n,u,g.
\end{equation}
Note that (\ref{subpro3-1-2}) is convex, which can be readily solved using convex optimization tools. An iterative algorithm can be proposed to solve (\ref{subpro3-1-1}) based on the solutions to (\ref{subpro3-1-2}) and (\ref{vp}), which is summarized in Algorithm 5. Then, to show the convergence of Algorithm 5, we can derive the Lagrangian dual function of (\ref{subpro3-1-2}) as
\begin{align} \label{Lagtau}
\nonumber & L(\mathbf{P}^{j}, \mathbf{v}^{j - 1}, \tau^{j}, \bm{\psi}, \bm{\nu}, \bm{\xi}, \bm{\theta}) =  \tau^{j} \\
\nonumber & + \sum_{n = 1}^{N} \sum_{u = 1}^{U_n} \psi_{n, u}(\sum_{g = 1}^{G} x^{r}_{n, u, g} T^{r - 1}_{n}  R_{a}(\mathbf{P}^{j}_{n, g}, v^{j - 1}_{n, u, g}) - \tau^{j}) \\
\nonumber & + \sum_{n = 1}^{N} \sum_{i = 1}^{N_s} \nu_{n, i} (\epsilon_p - \sum_{u = 1}^{U_n} \sum_{k = 1}^{K} \sum_{g = 1}^{G}   x^{r}_{n, u, g} y_{n, i, g} \tilde{l}_{n, i, g, k}^2 p^{j}_{n, g, k}) \\
\nonumber & + \sum_{k = 1}^{K} \xi_{k} ( E^{com}_{k} - \sum_{n = 1}^{N} \sum_{u = 1}^{U_n} \sum_{g = 1}^{G}  x^{r}_{n, u, g} p^{j}_{n, g, k} T^{r - 1}_n) \\
 & + \sum_{n = 1}^{N} \sum_{k = 1}^{K} \theta_{n, k}(p_{max} - \sum_{u = 1}^{U_n} \sum_{g = 1}^{G}  x^{r}_{n, u, g} p^{j}_{n, g, k}).
\end{align}
The function in (\ref{Lagtau}) is concave with respect to $\mathbf{P}^{j}$ and $\tau^j$, while convex with respect to $\mathbf{v}^{j - 1}$. Hence, Algorithm 5 is to find the saddle point of (\ref{Lagtau}), which is guaranteed to converge according to Theorem 1. Besides, the acquired saddle point is an extreme point of (\ref{Lagtau}) when $\mathbf{v}^{j - 1}$ is constant, so that $\tau^{j}$ is non-decreasing along with iterations according to [36, Theorem 2]. As a result, a locally optimal solution to (\ref{subpro3-1-1}) can be obtained by Algorithm~5.
}
 \addtolength{\topmargin}{0.05in}
 \begin{algorithm}[t]
\caption{Algorithm to solve (\ref{subpro3-1-1})}
\hspace*{0.02in} {\bf Input:}
$\left\{ E^{com}_{k} \ \ \forall k \right\}$, $p_{max}$, $\epsilon_{p}$.
\begin{algorithmic}[1]
\State Initialization: $\epsilon_{0} = 1 \times 10^{-3}$, $j = 1$, $\mathbf{P}^{0} = \mathbf{0}$, $\mathbf{v}^{0} = \mathbf{0}$, $\tau^0 = 0$;
\State Solve (\ref{subpro3-1-2}), denoting the solution as $(\mathbf{P}^{*}, \tau^*)$, set $\mathbf{P}^{1} = \mathbf{P}^{*}$, $\tau^1 = \tau^*$;
    \While{$|1 - \frac{ \tau^{j - 1}}{\tau^{j}}| > \epsilon_{0}$}
	\State Solve (\ref{vp}), denoting the solution as $\mathbf{v}^{*}$, set $\mathbf{v}^{j} = \mathbf{v}^{*}$;
	\State $j = j + 1$;
        \State Solve (\ref{subpro3-1-2}), denoting the solution as $(\mathbf{P}^{*}, \tau^*)$, set $\mathbf{P}^{j} = \mathbf{P}^{*}$, $\tau^j = \tau^*$;
    \EndWhile
\end{algorithmic}
\hspace*{0.02in} {\bf Output:}
 $\mathbf{P}^{j}$, $\tau^{j}$. \\
\end{algorithm}
}
\par{
Based on the solutions to (\ref{subpro3-x})--(\ref{subpro3-2}), (\ref{pro3}) can be iteratively solved. The steps of this method is recorded in Algorithm 6. Similar to Algorithm 3, the convergence of Algorithm 6 is guaranteed, and at least a locally optimal solution can be derived. 
}
\addtolength{\topmargin}{0.05in}
 \begin{algorithm}[t]
\caption{Minimum data transmission efficiency maximization algorithm}
\hspace*{0.02in} {\bf Input:}
$T_{total}$, $T_{max}$, $\left\{ E^{com}_{k} \ \ \forall k \right\}$, $p_{max}$, $\epsilon_{p}$.
\begin{algorithmic}[1]
\State Initialization: $\epsilon_{0} = 1 \times 10^{-2}$, $r = 1$, $\mathbf{T}^{0} = (T_{total} / N) \mathbf{1}$, $\mathbf{P}^{0} = \mathbf{0}$;
\State Solve (\ref{subpro3-x}), denoting the solution as $\mathbf{x}^{*}$, set $\mathbf{x}^{1} = \mathbf{x}^{*}$;
\State Solve (\ref{subpro3-1}), denoting the solution as $\mathbf{P}^{*}$, set $\mathbf{P}^{1} = \mathbf{P}^{*}$;
\State Solve (\ref{subpro3-2}), denoting the solution as $\mathbf{T}^{*}$, set $\mathbf{T}^{1} = \mathbf{T}^{*}$;
    \While{$|1 - \frac{ \min_{n, u}\sum_{g = 1}^{G} x^{r - 1}_{n, u, g} T^{r - 1}_{n}  R_{a}(\mathbf{P}^{r - 1}_{n, g}, w^{r - 1}_{n, u, g})}{ \min_{n, u} \sum_{g = 1}^{G} x^{r}_{n, u, g} T^{r}_{n}  R_{a}(\mathbf{P}^{r}_{n, g}, w^{r}_{n, u, g})}| > \epsilon_{0}$}
		 \State $r = r + 1$;
        \State Solve (\ref{subpro3-x}), denoting the solution as $\mathbf{x}^{*}$, set $\mathbf{x}^{r} = \mathbf{x}^{*}$;
\State Solve (\ref{subpro3-1}), denoting the solution as $\mathbf{P}^{*}$, set $\mathbf{P}^{r} = \mathbf{P}^{*}$;
\State Solve (\ref{subpro3-2}), denoting the solution as $\mathbf{T}^{*}$, set $\mathbf{T}^{r} = \mathbf{T}^{*}$;
    \EndWhile
\end{algorithmic}
\hspace*{0.02in} {\bf Output:}
 $\mathbf{x}^{r}$, $\mathbf{P}^{r}$, $\mathbf{T}^{r}$. \\
\end{algorithm}
\begin{remark}
\textnormal{Observing the methods proposed in Section \uppercase\expandafter{\romannumeral3} and Section \uppercase\expandafter{\romannumeral4}, we can find some similarities and differences. On one hand, both optimization problems are solved in an iterative way, because both of them focus on jointly allocating subchannels, transmit power and hovering times. On the other hand, time-sharing relaxation method can not be used to solve the max-min subchannel allocation subproblem, because subchannels can not be allocated to users in worst conditions due to the inadequate design of relaxation. }
\end{remark}
\section{Simulation Results and Discussions}
\par{ 
In this section, we present simulation results to evaluate the proposed algorithms. We consider a CSUN with a satellite and $K = 6$ UAVs in UAV swarm. We set the number of satellite users as $N_{s} = 10$, the total number of UAV users as $N_{U} = 200$, which are divided into $N = 20$ user groups with $10$ UAV users each. All UAV users are equipped with $M = 6$ antennas. We assume that the CSUN works at $5.8$ GHz with $G = 16$ subchannels.\footnote{In practice, more users can be served by using the proposed scheme, because there are more available subchannels. For example, if the bandwidth is $20$ MHz and the subcarrier spacing is $15$ kHz, there are at least $1200$ available subchannels. In this case, over $15000$ users can be served.} For the UAV channel, we generate the large-scale CSI based on real channel environment \cite{ITU525,ITU676} using a simulation software named as Visualyze 7. The locations of UAVs, UAV users and satellite users, as well as the subchannels of satellite users, are generated by this software, and the noise power is set as $\sigma^{2} = -107$ dBm. We set the interference temperature threshold as $\epsilon_{p} = -77$ dBm, parameters of practical constraints are set as $p_{max} = 300$ mW, $T_{total} = 100$ s, and $T_{max} = 7.5$ s for more flexible time scheduling. For simplicity, we assume that each UAV has the same $E^{com}_{k}$ for $\forall k$. The sum of $E^{com}_{k}$ is denoted as $E_{total}$, which is set as $E_{total} = 30$~J.
}
\begin{figure}[t] 
\centering
\includegraphics[width=3.4in]{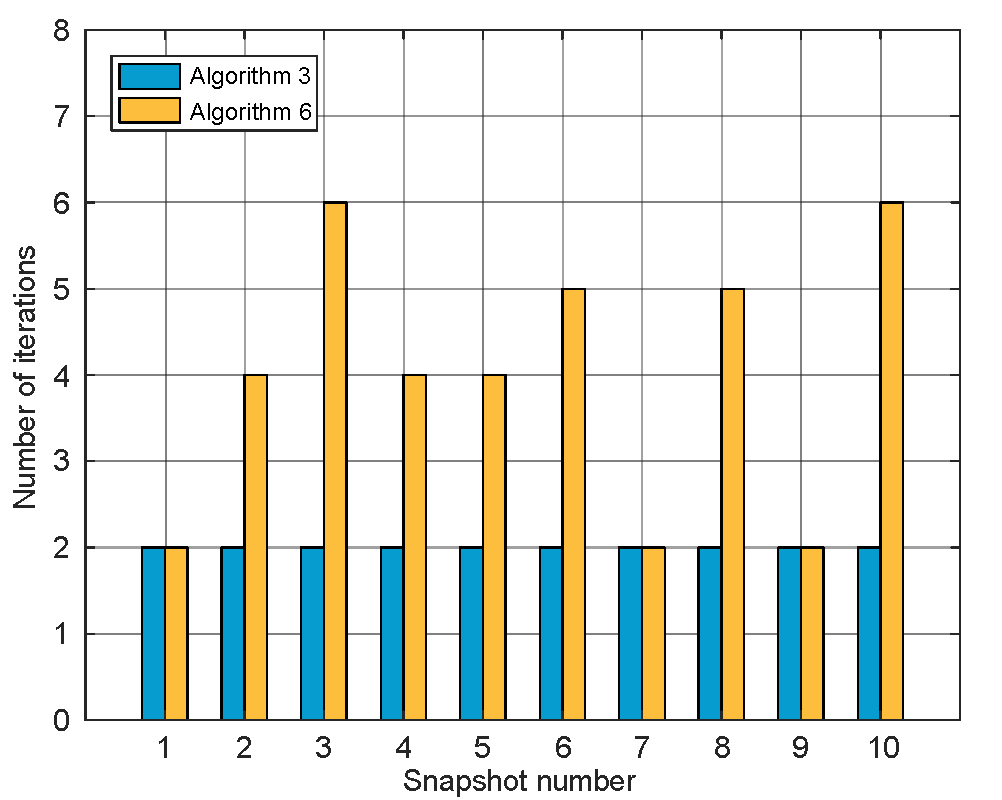}
\caption{ Convergence performance of the proposed algorithms.}
\label{Fig2}
\end{figure}
\par{
{We firstly verify the convergence performance of the proposed algorithms. In Fig.~\ref{Fig2}, $10$ snapshots with different user locations are evaluated. For Algorithm 3, it only needs $2$ iterations to converge. The reason is that the best subchannel can be selected at the first iteration, and given subchannel allocation the problem becomes convex with respect to transmit power and hovering time. For Algorithm 6, the number of iterations needed is no more than $6$. These results further indicate that these algorithms have good potentials in being applied to CSUNs in practice.
}
\begin{figure}[t] 
\centering
\includegraphics[width=3.5in]{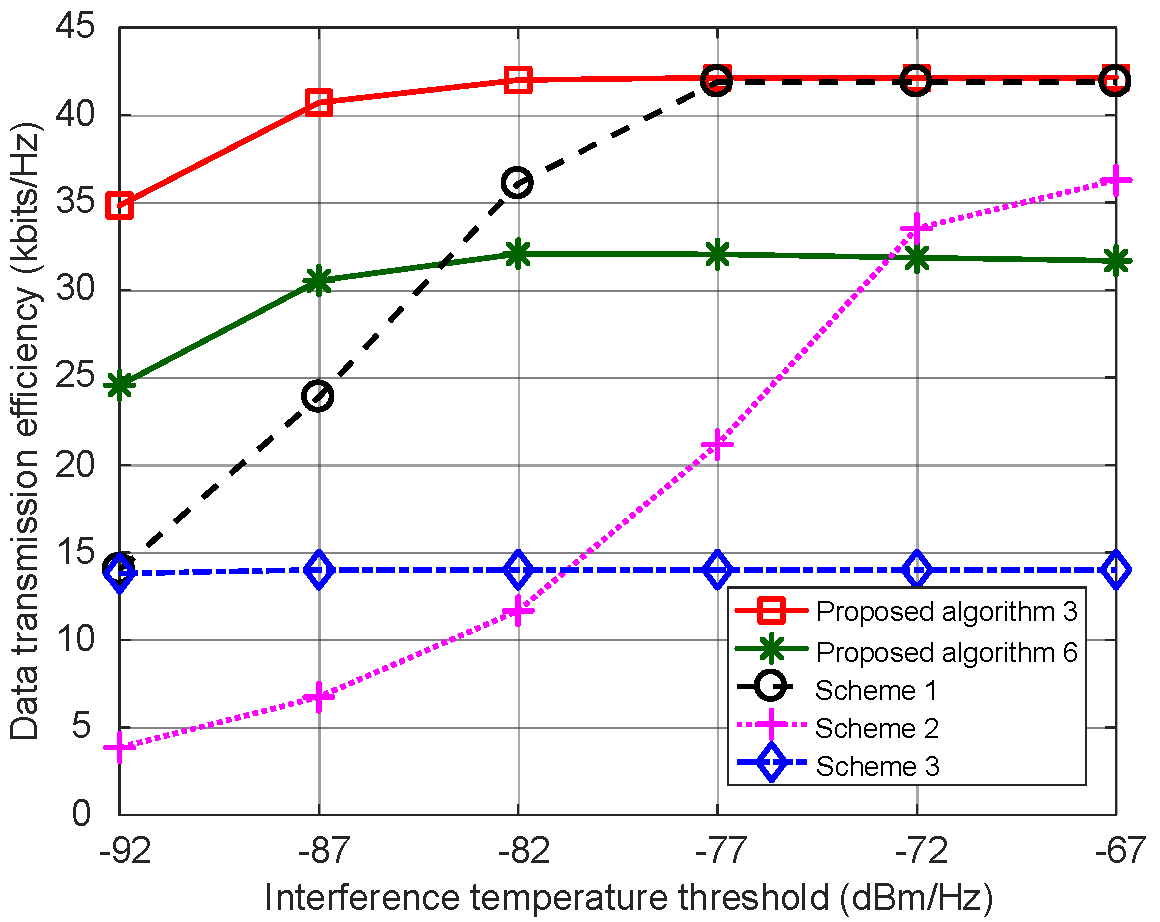}
\caption{Comparison of different algorithms considering the overall data transmission efficiency.}
\label{Fig3}
\end{figure}
\begin{figure}[t] 
\centering
\includegraphics[width=3.5in]{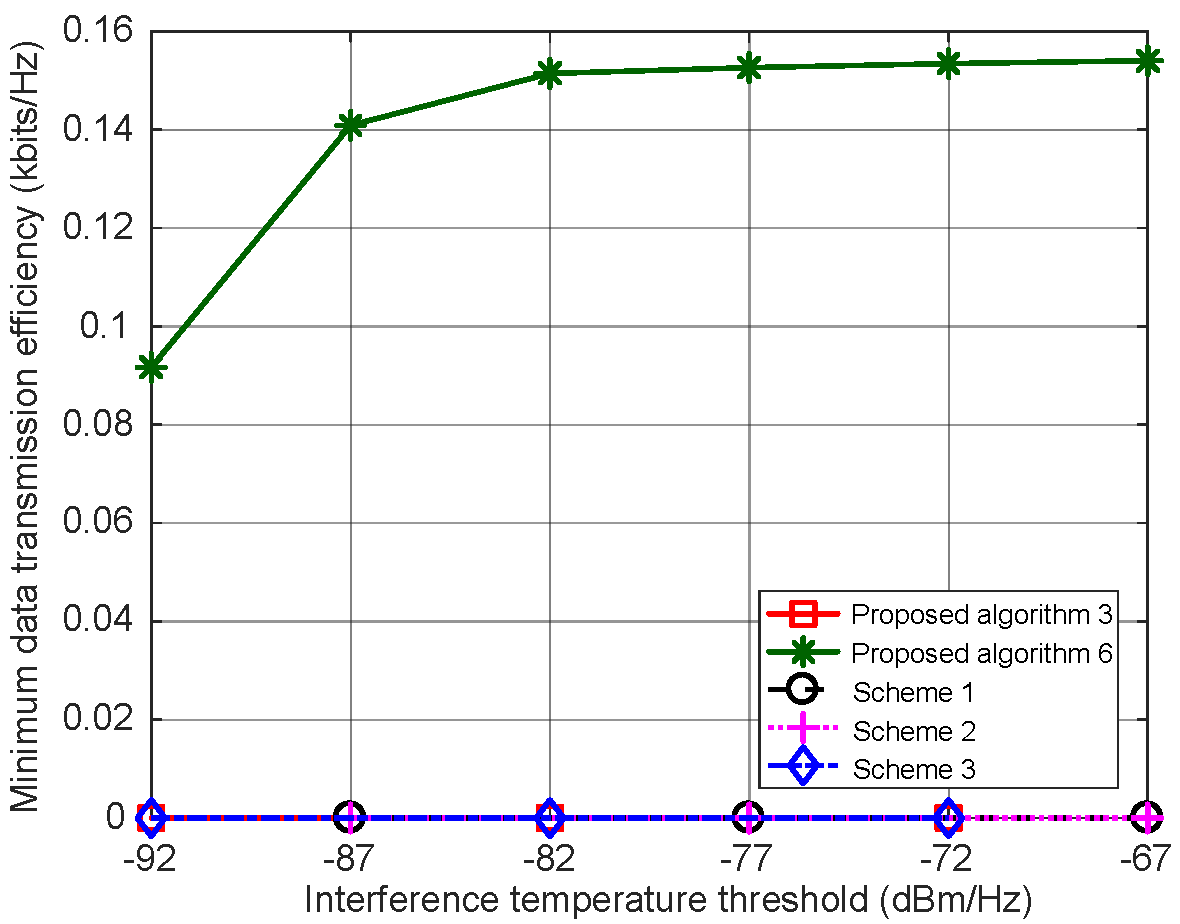}
\caption{Comparison of different algorithms regarding the minimum data transmission efficiency.}
\label{Fig4}
\end{figure}
\par{
Then, we compare the performances of the proposed algorithms with other algorithms. Because the formulated problems are non-convex, it is too time-consuming to search the globally optimal solution using brute force methods, even though we have cut down the number of subchannels for ease of simulations. Thereby, we compare the proposed algorithms with the following state-of-the-art methods.
\begin{itemize}
\item {\em Scheme 1:} Allocating the subchannels based on path loss using the cellular-based subchannel allocation method in \cite{Shen2005}, then using the power allocation algorithm and hovering time scheduling algorithm in \cite{Liu2019}.
\item {\em Scheme 2:} Allocating the subchannels based on path loss using the cellular-based subchannel allocation method in \cite{Shen2005}, then using the power allocation algorithm and hovering time scheduling algorithm in \cite{Hua2019}.
\item {\em Scheme 3:} Allocating the subchannels based on path loss using the cellular-based subchannel allocation method in \cite{Shen2005}, then equally allocating the transmit power and hovering times to all users.
\end{itemize}
Besides, for the power allocation algorithms in Scheme 1 and Scheme 3 which did not consider interference temperature constraints, the transmit power is divided by a large constant to satisfy these constraints. 
}
\par{In Fig.~\ref{Fig3}, we evaluate the performances of different algorithms in terms of data transmission efficiency with varying interference temperature thresholds, which can demonstrate the performance gain when the enlarged time scale of optimization is regarded. We can observe that when the interference temperature threshold increases, Scheme~1 approaches Algorithm~3. The reason is that although interference temperature constraints were not considered by Scheme~1, these constraints will be negligible when the interference temperature threshold is high. Moreover, we can find that with low interference temperature threshold, Scheme~2, in which the algorithm in \cite{Hua2019} did not use large-scale CSI, performs worse than Scheme~3. This result indicates that the underestimation of interference temperature can seriously affect the network efficiency. Besides, the overall data transmission efficiency can be to some extent improved by Algorithm~6, showing that a certain amount of network efficiency can be guaranteed when optimizing the user fairness. Furthermore, as shown in Fig.~\ref{Fig4}, we can observe that Algorithm 6 can effectively improve user fairness compared with other schemes. }
\begin{figure}[t]
\centering
\includegraphics[width= 3.5in]{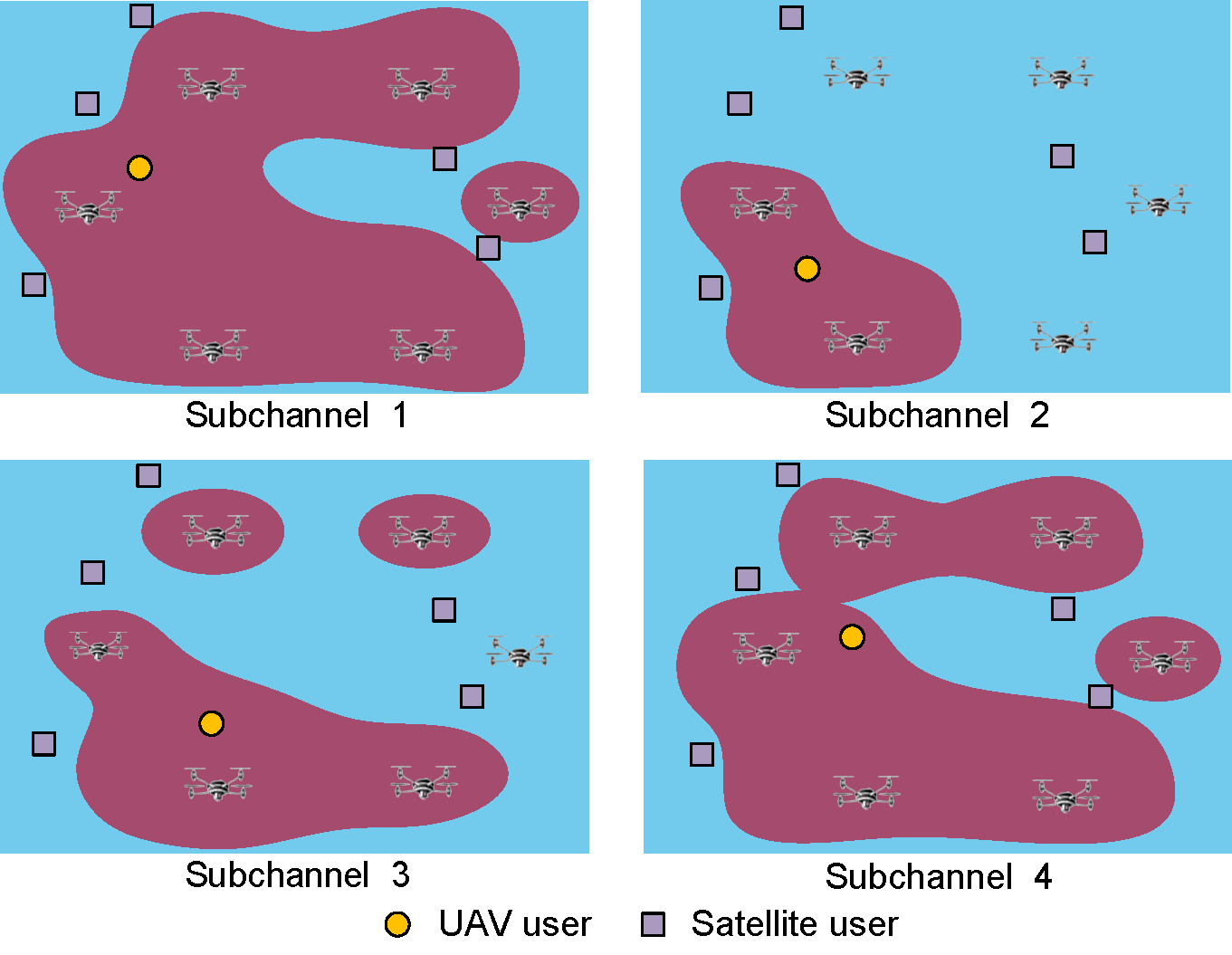}
\caption{Illustration of cell-free coverage areas in one time slot at different subchannels for CSUNs. }
\label{Fig5}
\end{figure}
\par{In Fig.~\ref{Fig5}, we demonstrate how the coverage of CSUN can be optimized by Algorithm 6. To make the figure more clear, we consider a simple scenario, where $4$ UAV users are served by UAVs in one time slot, $5$ satellite users are served by satellites, and the interference temperature threshold is set as $-92$ dBm. When the user can receive the signal whose power is larger than $-92$ dBm, this user is regarded to be successfully served. Following this, we can acquire the coverage areas when different subchannels are used. As shown in Fig.~\ref{Fig5}, the users in plotted regions can be served by UAVs. We can observe that the shapes of coverage areas are irregular, and both the shapes and the ranges of coverage areas vary with the change of subchannels. These results imply that cell-free CSUN can be efficiently established by using the proposed methods.}
\begin{figure}[t]
\centering
\includegraphics[width=3.5in]{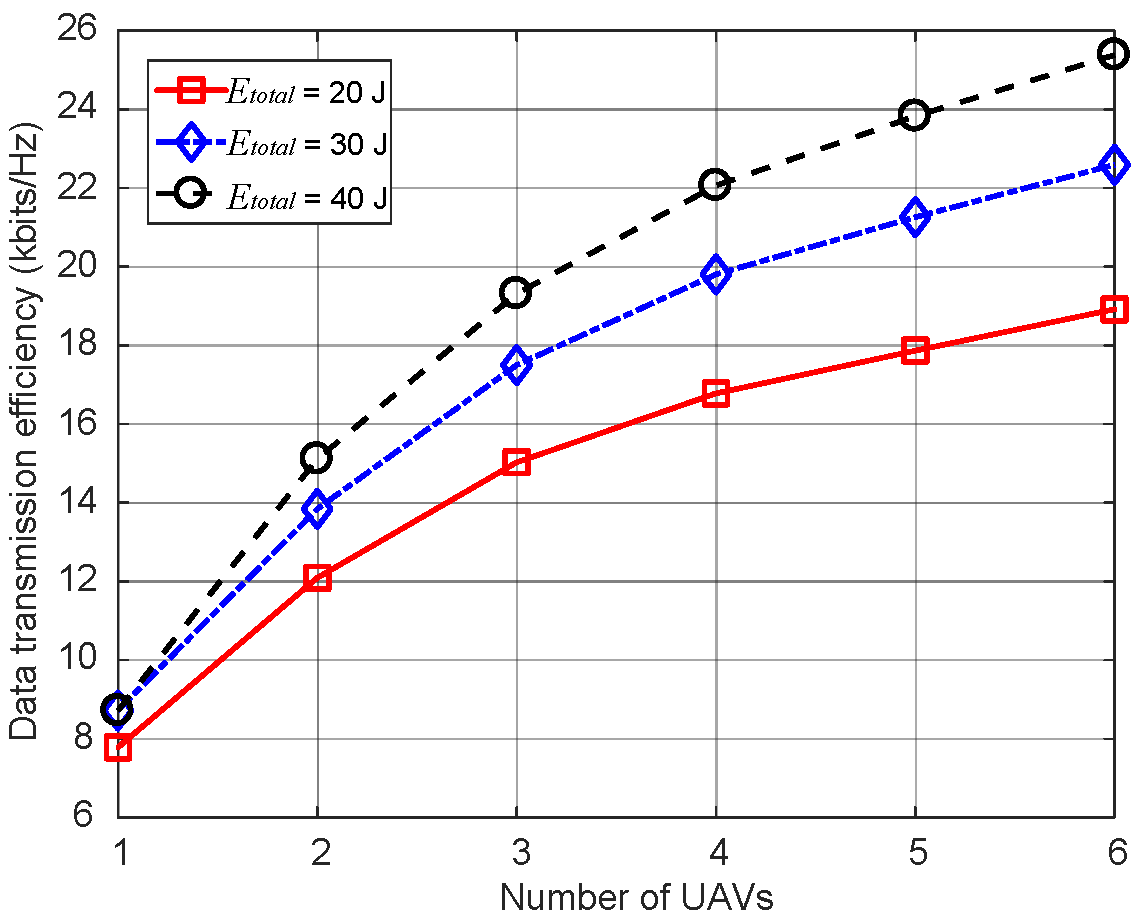}
\caption{The relationship between the overall data transmission efficiency and the number of UAVs, where the proposed Algorithm 3 is used.}
\label{Fig6}
\end{figure}
\begin{figure}[t]
\centering
\includegraphics[width=3.5in]{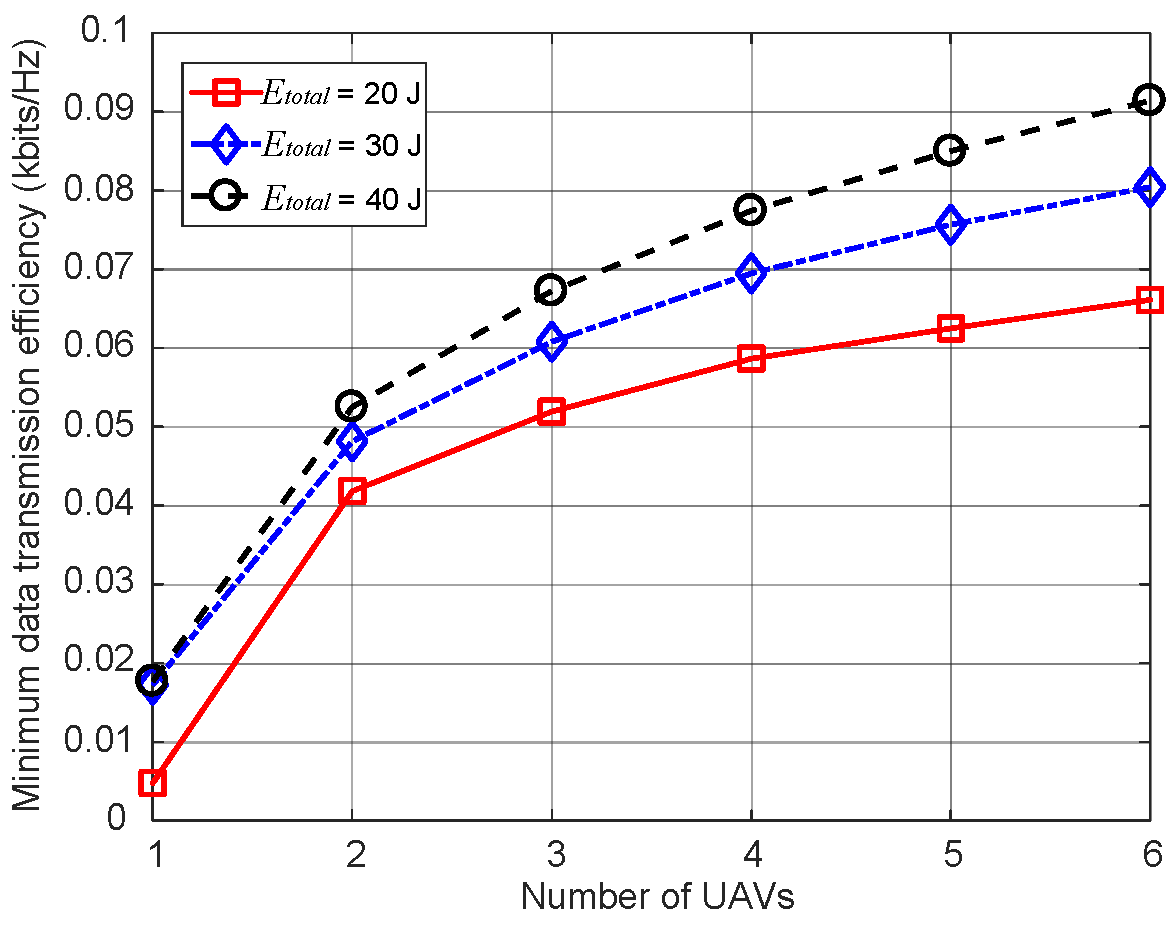}
\caption{The relationship between the minimum data transmission efficiency and the number of UAVs, where the proposed Algorithm 6 is used.}
\label{Fig7}
\end{figure}
\par{
Furthermore, we concentrate on analyzing the relationship between the size of UAV swarm and the performances of proposed algorithms in Fig.~\ref{Fig6} and Fig.~\ref{Fig7}. As shown by the curves, both the overall data transmission efficiency and the minimum data transmission efficiency can be improved by increasing the number of UAVs. One reason is that a higher diversity gain can be obtained with more UAVs in a swarm. Moreover, the coordination among multiple UAVs is more flexible when the size of UAV swarm is larger. Such phenomenon indicates that the use of coordinated multiple UAVs is an efficient way to cope with the varying practical channel environment in a large time scale. We can further observe that a better performance is achieved by both algorithms with higher communication energy. These results imply that the limited on-board energy of UAV swarm is a dominant bottleneck for cell-free CSUNs. 
}
\begin{figure}[t]
\centering
\includegraphics[width=3.5in]{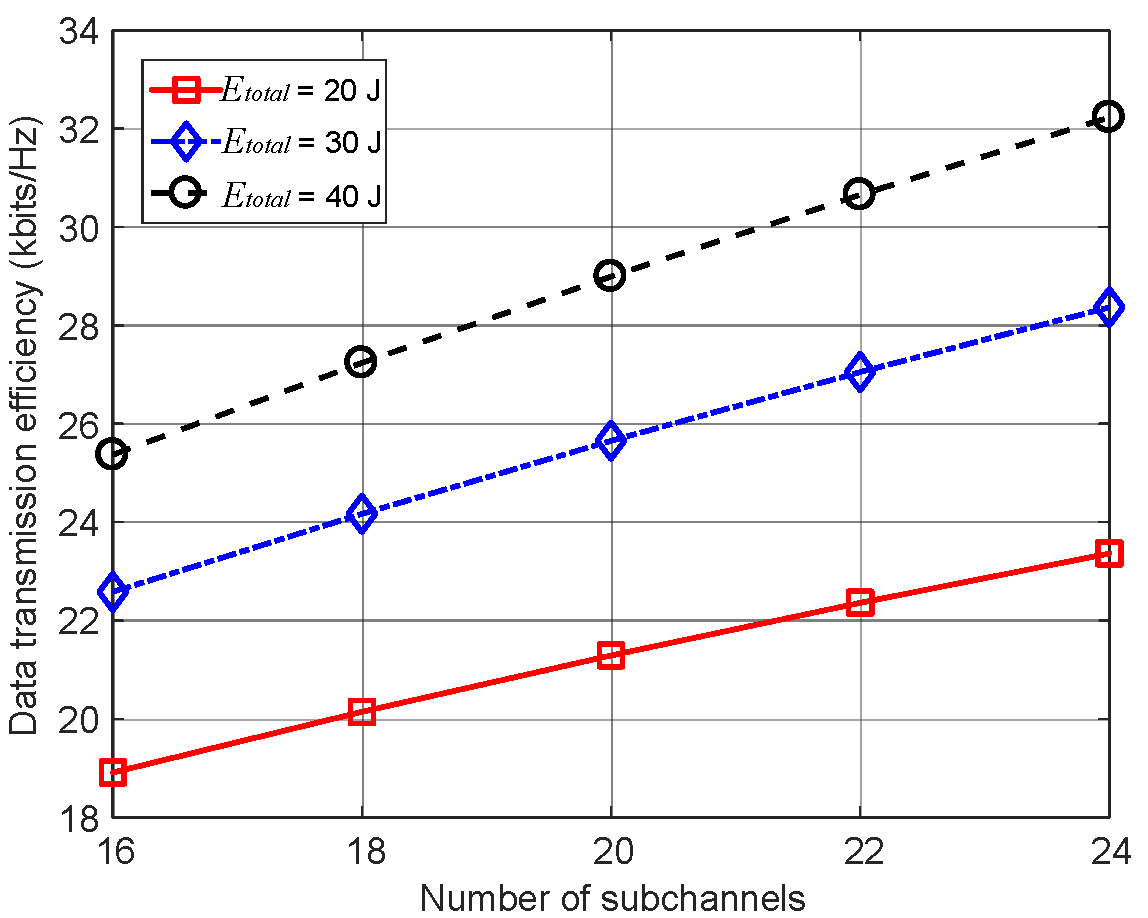}
\caption{The relationship between the overall data transmission efficiency and the number of subchannels, where the proposed Algorithm 3 is used.}
\label{Fig8}
\end{figure}
\begin{figure}[t]
\centering
\includegraphics[width=3.5in]{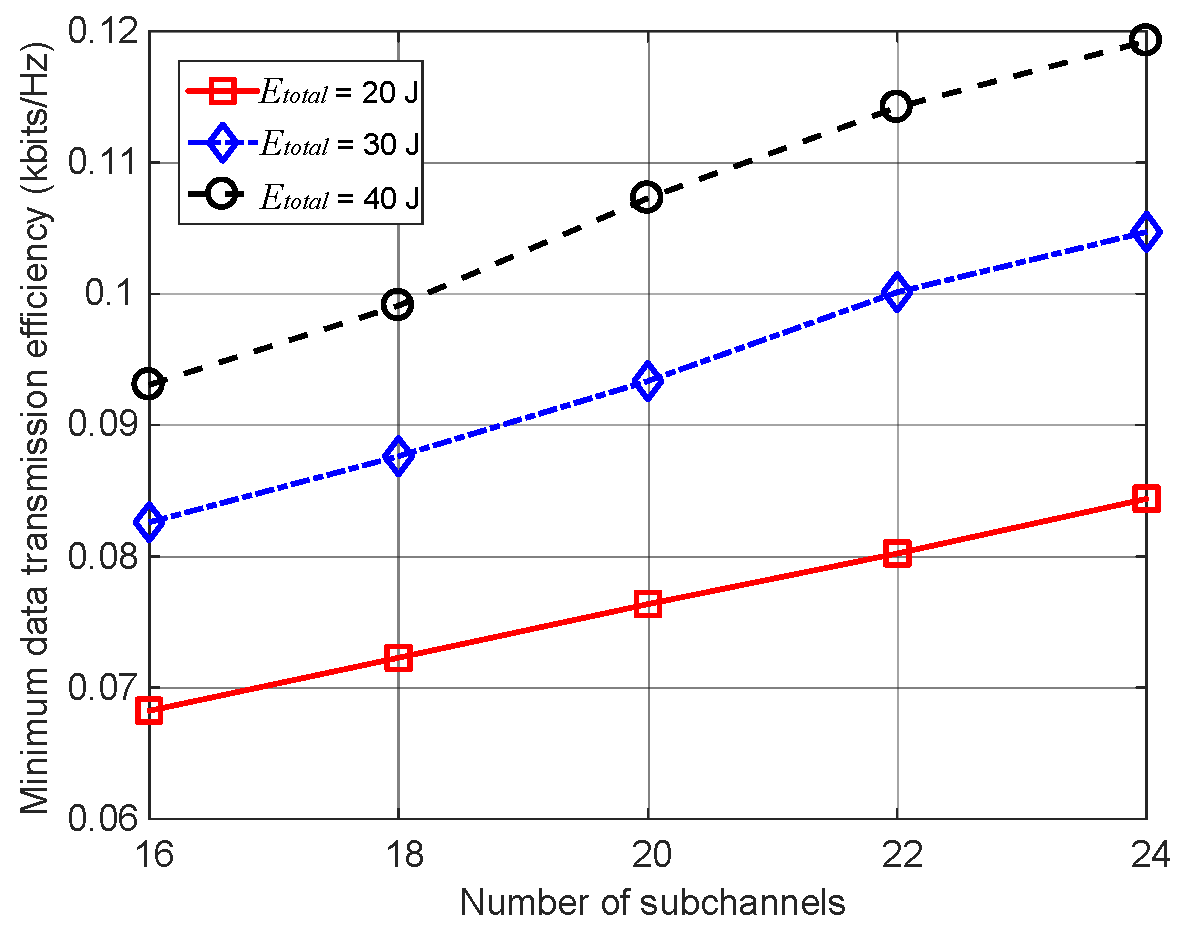}
\caption{The relationship between the minimum data transmission efficiency and the number of subchannels, where the proposed Algorithm 6 is used.}
\label{Fig9}
\end{figure}
\par{
In Fig.~\ref{Fig8} and Fig.~\ref{Fig9}, we evaluate the relationship between the number of subchannels and the performances of the proposed algorithms. We can observe that better performance is achieved when more subchannels are used for both algorithms. Besides, the curves in Fig.~\ref{Fig9} demonstrate that the performance gain of improving communication energy fluctuates for different number of subchannels. This phenomenon emerges because a locally optimal solution is derived by Algorithm 6, which implies that the number of available subchannels should be appropriately designed for more efficient use of resources in cell-free CSUNs. 
}
\section{Conclusions}
\par{
In this paper, we have investigated multi-domain resource allocation for cell-free IoT-oriented CSUNs, to support massive access for IoT devices outside terrestrial cellular networks. We have proposed a process-oriented optimization framework, where the whole flight process of UAVs was optimized only using slowly-varying large-scale CSI. We have formulated a data transmission efficiency maximization problem and a minimum data transmission efficiency maximization problem under the process-oriented framework to improve network efficiency with guaranteed user fairness. After the optimization problems have been solved using the time-sharing relaxation and feasible region relaxation methods, the subchannels, transmit power and hovering times are jointly allocated in an iterative way. Simulation results have demonstrated that it is beneficial to use the proposed methods. Moreover, the cell-free coverage pattern has been observed by using proposed algorithms in the simulation, which indicates a promising way to efficiently support massive access for wide-area IoT devices in the upcoming 6G era.
}
\begin{appendices}
\section{Proof of Theorem 2}
\par{
 Assuming that $\mathbf{P}^{r - 1}$ and $\mathbf{x}^{r - 1}$ have been obtained at the $(r-1)$-th step. For an all-zero vector $\mathbf{x}$, the constraints in (\ref{subpro3-x1c})--(\ref{subpro3-x1e}) are naturally satisfied. Otherwise, for any non-zero $\mathbf{x}$ which satisfies (\ref{subpro3-x1b}), (\ref{subpro3-x1f}) and (\ref{subpro3-x1g}) with any given $n^* \in \{1,...,N\}$, $g^* \in \{1,...,G\}$ and $k^* \in \{1,...,K\}$, we have
\begin{equation}
\sum_{u = 1}^{U_{n^{*}}} x_{n^*, u, g^*} p^{r - 1}_{n^*, g^* ,k^*} = \sum_{u = 1}^{U_{n^{*}}} x^{r - 1}_{n^*, u, g^*} p^{r - 1}_{n^*, g^*,k^*} \tag{A.1} \label{xp}
\end{equation}
because only one element in $\left\{x^{r - 1}_{n^*, u, g^*}, u\in\{1,..., U_{n^*}\} \right\}$ equals to $1$ according to (\ref{subpro3-x1f}) and (\ref{subpro3-x1g}), which is also correct for $\mathbf{x}$. Hence, we have
\begin{align}
 \nonumber & \sum_{u = 1}^{U_n}\sum_{k = 1}^{K} \sum_{g = 1}^{G}x_{n,u,g} y_{n, i, g} p^{r - 1}_{n, g, k}   \tilde{l}_{n, i, g, k}^2  \\
\nonumber & = \sum_{u = 1}^{U_n}\sum_{k = 1}^{K} \sum_{g = 1}^{G}x^{r - 1}_{n,u,g} y_{n, i, g} p^{r - 1}_{n, g, k}   \tilde{l}_{n, i, g, k}^2  \\
& \leq \epsilon_p \ \ \forall n,i \tag{A.2} \label{a1}
\end{align}
\begin{align}
\nonumber &  \sum_{n = 1}^{N} \sum_{u = 1}^{U_n}  \sum_{g = 1}^{G}  x_{n,u,g} p^{r - 1}_{n, g, k} T^{r - 1}_n  \\
 & = \sum_{n = 1}^{N} \sum_{u = 1}^{U_n}  \sum_{g = 1}^{G}  x^{r-1}_{n,u,g} p^{r - 1}_{n, g, k} T^{r - 1}_n  \leq E^{com}_{k} \ \ \forall k \tag{A.3} \label{a2} 
\end{align}
\begin{align}
\nonumber &\sum_{u = 1}^{U_n} \sum_{g = 1}^{G}  x_{n,u,g} p^{r - 1}_{n, g, k} \\
& = \sum_{u = 1}^{U_n} \sum_{g = 1}^{G}  x^{r - 1}_{n,u,g} p^{r - 1}_{n, g, k} \leq p_{max} \ \ \forall n,k \tag{A.4} \label{a3}
\end{align}
based on (\ref{xp}). Observing (\ref{a1})--(\ref{a3}), we can find that for any $\mathbf{x}$ that satisfies (\ref{subpro3-x1b}), (\ref{subpro3-x1f}) and (\ref{subpro3-x1g}), (\ref{subpro3-x1c})--(\ref{subpro3-x1e}) are also satisfied. As a result, (\ref{subpro3-x1c})--(\ref{subpro3-x1e}) actually have no influence on (\ref{subpro3-x1}), which means the conclusion of Theorem 2 is given.
}
\section{Proof of Theorem 3}
\par{
Substituting $\mathbf{v}^{r} = \{ v^{r}_{n,u,g} \ \ \forall n,u,g\}$ into (\ref{subpro3-1-1b}) and (\ref{subpro3-1-1f}), we have
\begin{align}
& \sum_{g = 1}^{G} x^{r}_{n, u, g} T^{r - 1}_{n}  R_{a}(\mathbf{P}^{r}_{n, g}, v^{r}_{n, u, g}) \geq \tau \ \ \forall n,g \tag{B.1} \label{c1}\\
  &  e^{v^{r}_{n,u,g}} = 1 + \sum_{k = 1}^{K} \frac{l_{n,u,g,k}^{2} p^{r}_{n, g, k}}{\sigma^{2} + Ml_{n,u,g,k}^{2} p^{r}_{n, g, k} e^{-v^{r}_{n,u,g}}}  \ \ \forall n,u,g. \tag{B.2} \label{c2} 
\end{align}
According to \cite{Liu2019}, if $\mathbf{v}^{r*}$ satisfies (\ref{c1}) and (\ref{c2}), we have
\begin{equation}
R_{a}(\mathbf{P}^{r}_{n, g}, v^{r}_{n, u, g}) \geq R_{a}(\mathbf{P}^{r}_{n, g}, v^{r*}_{n, u, g}) \geq \tau \ \ \forall v^{r}_{n, u, g} \geq 0 \tag{B.3} \label{c3}
\end{equation}
because the minimum value of $R_{a}(\mathbf{P}^{r}_{n, g}, v^{r}_{n, u, g})$ is achieved by $R_{a}(\mathbf{P}^{r}_{n, g}, v^{r*}_{n, u, g})$. Thus, we have
\begin{align}
& \sum_{g = 1}^{G} x^{r}_{n, u, g} T^{r - 1}_{n}  R_{a}(\mathbf{P}^{r}_{n,  g}, v^{r}_{n, u, g}) \geq \tau \ \ \forall n,g \tag{B.4} \label{c4} \\  
& v^{r}_{n, u, g} \geq 0 \ \ \forall n,u,g. \tag{B.5} \label{c5}
\end{align}
On the contrary, if (\ref{c4}) and (\ref{c5}) are satisfied, we can also have (\ref{c1}) and (\ref{c2}), because (\ref{c4}) and (\ref{c5}) are more general conditions. Hence, (\ref{c1}) and (\ref{c2}) are equivalent to (\ref{c4}) and (\ref{c5}), which gives the conclusion of Theorem~3.
}
\end{appendices}
\bibliographystyle{ieeetr}

\begin{thebibliography}{10}
\bibitem{Liu2020WOCC}	
C. Liu, W. Feng, Y. Chen, C.-X. Wang, X. Li, and N. Ge, ``Process-oriented optimization for Beyond 5G cognitive satellite-UAV networks (invited paper),'' in {\em Proc. IEEE WOCC}, Newark, NJ, USA, 2020, pp.~1-6.
\bibitem{Jia2020}
R. Jia, X. Chen, Q. Qi, and H. Lin, ``Massive beam-division multiple access for B5G cellular Internet of Things,'' {\em IEEE Internet Things J.}, vol.~7, no.~3, pp.~2386-2396, Mar.~2020.
\bibitem{Qi2020}
Q. Qi, X. Chen, and D. W. K. Ng, ``Robust beamforming for NOMA-based cellular massive IoT with SWIPT,'' { \em IEEE Trans. Signal Process.}, vol.~68, pp.~211-224, 2020.
\bibitem{Ikpehai2019}
A. Ikpehai {\em et al.}, ``Low-Power Wide Area Network technologies for Internet-of-Things: a comparative review,'' {\em IEEE Internet Things J.}, vol.~6, no.~2, pp.~2225-2240, Apr.~2019.
\bibitem{Zhen2019}
L. Zhen {\em et al.}, ``Optimal preamble design in spatial group-based random access for satellite-M2M communications,'' {\em IEEE Wireless Commun. Lett.}, vol.~8, no.~3, pp.~953-956, Jun.~2019.
\bibitem{Yang2018}
Q. Yang and S. Yoo, ``Optimal UAV path planning: sensing data acquisition over IoT sensor networks using multi-objective bio-inspired algorithms,'' {\em IEEE Access}, vol.~6, pp. 13671-13684, 2018.
\bibitem{Baek2019}
J. Baek, S. I. Han, and Y. Han, ``Energy-efficient UAV routing for wireless sensor networks,'' {\em IEEE Trans. Veh. Tech.}, vol.~69, no.~2, pp.~1741-1750, Feb.~2020.
\bibitem{Lyu2018}
J. Lyu, Y. Zeng, and R. Zhang, ``UAV-aided offloading for cellular hotspot,'' {\em IEEE Trans. Wireless Commun.}, vol.~17, no.~6, pp. 3988-4001, Jun.~2018.
\bibitem{Pandian2015}
M. B. Pandian, M. L. Sichitiu, and H. Dai, ``Optimal resource allocation in random access cooperative cognitive radio networks,'' {\em IEEE Trans. Mobile Comput.}, vol.~14, no.~6, pp.~1245-1258, Jun.~2015.
\bibitem{Raza2017}
U. Raza, P. Kulkarni, and M. Sooriyabandara, ``Low Power Wide Area Networks: an overview,'' {\em IEEE Commun. Surv. Tut.}, vol.~19, no.~2, pp.~855-873, Secondquart.~2017.
\bibitem{Popli2019}
S. Popli, R. K. Jha, and S. Jain, ``A survey on energy efficient Narrowband Internet of Things (NBIoT): architecture, application and challenges,'' {\em IEEE Access}, vol.~7, pp.~16739-16776, 2019.
\bibitem{Onireti2016}
O. Onireti, M. A. Imran, J. Qadir, and A. Sathiaseelan, ``Will 5G see its blind side? evolving 5G for universal Internet access,'' in {\em Proc. ACM GAIA Workshop}, Florianopolis, Brazil, Aug.~2016, pp.~1-6.
\bibitem{Centenaro2016}
M. Centenaro, L. Vangelista, A. Zanella, and M. Zorzi, ``Long-range communications in unlicensed bands: the rising stars in the IoT and smart city scenarios,'' {\em IEEE Wireless Commun.}, vol. 23, no. 5, pp. 60-67, Oct.~2016.
\bibitem{Sanctis2016}
M. De Sanctis, E. Cianca, G. Araniti, I. Bisio, and R. Prasad, ``Satellite communications supporting Internet of Remote Things,'' {\em IEEE Internet Things J.}, vol.~3, no.~1, pp.~113-123, Feb.~2016.
\bibitem{Wang2018}
Q. Wang, G. Ren, S. Gao, and K. Wu, ``A framework of Non-Orthogonal Slotted Aloha (NOSA) protocol for TDMA-based random multiple access in IoT-oriented satellite networks,'' {\em IEEE Access}, vol.~6, pp.~77542-77553, 2018.
\bibitem{Maleki2015}
S. Maleki {\em et al.}, ``Cognitive spectrum utilization in Ka band multibeam satellite communications,'' {\em IEEE Commun. Mag.}, vol.~53, no.~3, pp.~24-29, Mar.~2015.
\bibitem{Vazquez2018}
M. \'A. V\'azquez, L. Blanco, and A. I. P\'erez-Neira, ``Hybrid analog-digital transmit beamforming for spectrum sharing backhaul networks,'' {\em IEEE Trans. Signal Process.}, vol.~66, no.~9, pp.~2273-2285, May~2018.
\bibitem{Khan2012}
A. H. Khan, M. A. Imran, and B. G. Evans, ``Semi-adaptive beamforming for OFDM based hybrid terrestrial-satellite mobile system,'' {\em IEEE Trans. Wireless Commun.}, vol.~11, no.~10, pp.~3424-3433, Oct.~2012.
\bibitem{Liu2019Optimal}
C. Liu, W. Feng, Y. Chen, C.-X. Wang, and N. Ge, ``Optimal beamforming for hybrid satellite terrestrial networks with nonlinear PA and imperfect CSIT,'' {\em IEEE Wireless Commun. Lett.}, vol.~9, no.~3, pp.~276-280, Mar.~2020.
\bibitem{Feng2018UAV}
W. Feng, J. Wang, Y. Chen, X. Wang, N. Ge, and J. Lu, ``UAV-aided MIMO communications for 5G Internet of Things,'' {\em IEEE Internet Things J.}, vol.~6, no.~2, pp.~1731-1740, Apr.~2019.
\bibitem{Zeng2019}
Y. Zeng, Q. Wu, and R. Zhang, ``Accessing from the sky: a tutorial on UAV communications for 5G and beyond,'' {\em Proc. IEEE}, vol.~107, no.~12, pp.~2327-2375, Dec.~2019.
\bibitem{Salam2019}
T. Salam, W. U. Rehman, and X. Tao, ``Data aggregation in massive machine type communication: challenges and solutions,'' {\em IEEE Access}, vol. 7, pp. 41921-41946, 2019.
\bibitem{Hattab2020}
G. Hattab and D. Cabric, ``Energy-efficient massive IoT shared spectrum access over UAV-enabled cellular networks,'' {\em IEEE Trans. Commun.}, doi: 10.1109/TCOMM.2020.2998547.
\bibitem{Bushnaq2019}
O. M. Bushnaq, A. Celik, H. Elsawy, M. Alouini, and T. Y. Al-Naffouri, ``Aeronautical data aggregation and field estimation in IoT networks: hovering and traveling time dilemma of UAVs,'' {\em IEEE Trans. Wireless Commun.}, vol.~18, no.~10, pp.~4620-4635, Oct.~2019.
\bibitem{Wu2018}
Q. Wu, Y. Zeng, and R. Zhang, ``Joint trajectory and communication design for multi-UAV enabled wireless networks,'' {\em IEEE Trans. Wireless Commun.}, vol.~17, no.~3, pp.~2109-2121, Mar.~2018.
\bibitem{Khuwaja2019}
A. Khuwaja, G. Zheng, Y. Chen, and W. Feng, ``Optimum deployment of multiple UAVs for coverage area maximization in the presence of co-channel interference,'' {\em IEEE Access}, vol. 7, pp. 85203-85212, 2019.
\bibitem{Hua2019}
M. Hua, Y. Wang, M. Lin, C. Li, Y. Huang, and L. Yang, ``Joint CoMP transmission for UAV-aided cognitive satellite terrestrial networks,'' {\em IEEE Access}, vol.~7, pp.~14959-14968, 2019.
\bibitem{Wang2019}
T. Qi, W. Feng, and Y. Wang, ``Outage performance of non-orthogonal multiple access based unmanned aerial vehicles satellite networks,'' \emph{China Commun.}, vol. 15, no. 5, pp. 1-8, May 2018.
\bibitem{ITU525}
ITU-R Recommendation, ``Calculation of free-space attenuation,'' Int. Telecommun. Union, Geneva, Switzerland, ITU-R P. 525-4, 2019.
\bibitem{ITU676}
ITU-R Recommendation, ``Attenuation by atmospheric gases,'' Int. Telecommun. Union, Geneva, Switzerland, ITU-R P. 676-12, 2019.
\bibitem{Monwar2018}
M. Monwar, O. Semiari, and W. Saad, ``Optimized path planning for inspection by Unmanned Aerial Vehicles swarm with energy constraints,'' in {\em Proc. IEEE GLOBECOM}, Abu Dhabi, United Arab Emirates, 2018, pp.~1-6.
\bibitem{Liu2019}
C. Liu, W. Feng, J. Wang, Y. Chen, and N. Ge, ``Aerial small cells using coordinated multiple UAVs: an energy efficiency optimization perspective,'' {\em IEEE Access}, vol.~7, pp.~122838-122848, 2019.
\bibitem{Chen2018}
Y. Chen, W. Feng, and G. Zheng, ``Optimum placement of UAV as relays,'' {\em IEEE Commun. Lett.}, vol.~22, no.~2, pp.~248-251, Feb.~2018.
\bibitem{CXWang2018}
C.-X. Wang, J. Bian, J. Sun, W. Zhang, and M. Zhang, ``A survey of 5G channel measurements and models,'' {\em IEEE Commun. Surv. Tut.}, vol.~20, no.~4, pp.~3142-3168, Fourthquart.~2018.
\bibitem{Wangxx2019}
X. Wang, W. Feng, Y. Chen, and N. Ge, ``UAV swarm-enabled aerial CoMP: a physical layer security perspective,'' {\em IEEE Access}, vol.~7, pp.~120901-120916, 2019.
\bibitem{Feng2013}
W. Feng, Y. Wang, N. Ge, J. Lu, and J. Zhang, ``Virtual MIMO in multicell distributed antenna systems: coordinated transmissions with large-scale CSIT,'' {\em IEEE J. Sel. Areas Commun.}, vol.~31, no.~10, pp.~2067-2081, Oct.~2013.
\bibitem{Joung2014}
J. Joung, Y. K. Chia, and S. Sun, ``Energy-efficient, large-scale distributed antenna system (L-DAS) for multiple users,'' {\em IEEE J. Sel. Topics Signal Process.}, vol.~8, no.~5, pp.~954-965, Oct.~2014.
\bibitem{Boyd2004Convex}
S. Boyd and L. Vandenberghe, {\em Convex optimization}. Cambridge University Press, 2004.
\bibitem{Wei2019}
T. Wei, W. Feng, J. Wang, N. Ge, and J. Lu, ``Exploiting the shipping lane information for energy-efficient maritime communications,'' {\em IEEE Trans. Veh. Tech.}, vol.~68, no.~7, pp.~7204-7208, Jul.~2019.
\bibitem{Murray1980}
W. Murray and M. L. Overton, ``A projected Lagrangian algorithm for nonlinear minimax optimization,'' {\em SIAM J. Sci. Stat. Comput.}, vol.~1, no.~3, pp.~345-370, Sep.~1980.
\bibitem{Shen2005}
Z. Shen, J. G. Andrews, and B. L. Evans, ``Adaptive resource allocation in multiuser OFDM systems with proportional rate constraints,'' {\em IEEE Trans. Wireless Commun.}, vol.~4, no.~6, pp.~2726-2737, Nov.~2005.
\bibitem{Sun2017}
Y. Sun, P. Babu, and D. P. Palomar, ``Majorization-minimization algorithms in signal processing, communications, and machine learning,'' {\em IEEE Trans. Signal Process.}, vol.~65, no.~3, pp.~794-816, Feb.~2017.
\end{thebibliography}

\end{document}